%% file: main_arxiv.tex
\documentclass{ectj}
\usepackage{natbib}
\usepackage{amsmath,amssymb,amscd,amsfonts,amsbsy}
\usepackage{bm}
\usepackage{dsfont}
\usepackage{bbm}
\usepackage[inline]{enumitem}
\usepackage{multirow}
\usepackage{graphicx,graphics,rotating}
\usepackage[graphicx]{realboxes} 
\usepackage{epstopdf} 
\usepackage[para,online,flushleft]{threeparttable}	
\usepackage{booktabs,array}  
\usepackage{subfig}
\usepackage{url}

\usepackage{xr}

\makeatletter

\input{new_commands}

\makeatletter
\newcommand{\distas}[1]{\mathbin{\overset{#1}{\kern\z@\sim}}}%
\newsavebox{\mybox}\newsavebox{\mysim}
\newcommand{\distras}[1]{%
  \savebox{\mybox}{\hbox{\kern3pt$\scriptstyle#1$\kern3pt}}%
  \savebox{\mysim}{\hbox{$\sim$}}%
  \mathbin{\overset{#1}{\kern\z@\resizebox{\wd\mybox}{\ht\mysim}{$\sim$}}}%
}

\newtheorem{assumption}{Assumption}

\newtheorem{algorithm}{Algorithm}

\renewcommand{\thesection}{\arabic{section}}
\renewcommand{\theequation}{\arabic{section}.\arabic{equation}}

\newcounter{bean}

\received{August 2024}
  
\title[Double Machine Learning for Static Panel
Models with Fixed Effects]{Double Machine Learning for Static Panel
Models with Fixed Effects}

\author[P. S. Clarke and A. Polselli]{Paul S. Clarke$^{\dagger}$ and
                        Annalivia Polselli$^{\ddagger}$}

\address{$^{\dagger}$Institute for Social and Economic Research, University of Essex, Colchester CO4 3SQ, UK.}
\email{pclarke@essex.ac.uk}
\address{$^{\ddagger}$Institute for Analytics and Data Science,  University of Essex, Colchester CO4 3ZL, UK.}
\email{annalivia.polselli@essex.ac.uk}

\def\AmSTeX{$\cal A$\kern-.1667em\lower.5ex\hbox{$\cal M$}\kern-.125em
            $\cal S$-\TeX}

\begin{document}
  \begin{abstract}
Recent advances in causal inference have seen the development of methods which make use of the predictive power of machine learning algorithms. In this paper, we develop novel double machine learning (DML) procedures for panel data in which these algorithms are used to approximate high-dimensional and nonlinear nuisance functions of the covariates. Our new procedures are extensions of the well-known correlated random effects, within-group and first-difference estimators from linear to nonlinear panel models, specifically, \citet{robinson1988}’s partially linear regression model with fixed effects and unspecified nonlinear confounding. Our simulation study assesses the performance of these procedures using different machine learning algorithms. We use our procedures to re-estimate the impact of minimum wage on voting behaviour in the UK. From our results, we recommend the use of first-differencing because it imposes the fewest constraints on the distribution of the fixed effects, and an ensemble learning strategy to ensure optimum estimator accuracy.

  \end{abstract}

\input{01_introduction.tex}

\input{02_model_estimators}
\input{03a_estimation_inference}
\input{04_simulations.tex}
\input{05_application.tex}
\input{06_conclusion.tex}\section*{Acknowledgements}
The authors are grateful to  Andreas Alfons, Anna Baiardi, Thomas Cornelissen, Riccardo Di Francesco, Maria Grith, Omar Hussein, Damian Machlanski, Spyros Samothrakis, David Zentler-Munro, Wendun Wang, and the participants to the Annual MiSoC Workshop,  RSS International Conference 2023, RCEA-ICEEF 2024, IPDC 2024, the internal seminars in ISER and in the Econometric Institute at the Erasmus University Rotterdam for the helpful comments and suggestions. We also thank the editor Christoph Rothe and three anonymous referees, whose comments significantly improved the final version of this paper.
This research was funded by the UK Economic and Social Research Council award ES/S012486/1 (MiSoC).
The authors also acknowledge the use of the High Performance Computing Facility (Ceres) and its associated support services at the University of Essex in the completion of this work.

%

\bibliography{biblio}

\newpage
\clearpage
\renewcommand{\thesection}{S\arabic{section}}
\setcounter{section}{0}
\section*{Online Supplementary Information}
\input{888_proof_prop1}
\input{888_score}
\input{SI_mc_simulations.tex}
\input{SI_tuning.tex}
\input{SI_var_description}

\end{document}

%% file: new_commands.tex
\def\vector#1{\mbox{\boldmath{$#1$}}} 

\newcommand{\thetano}{\widehat{\theta}}

\newcommand{\btheta}{\bm{\theta}}

\newcommand{\x}{\mathbf{x}}
\newcommand{\Xit}{\mathbf{X}_{it}}

\newcommand{\Xii}{\mathbf{X}_{i}}
\newcommand{\bxi}{{\overline{X}}_{i}}

\newcommand{\Dit}{D_{it}}
\newcommand{\dit}{d_{it}}

\newcommand{\bdi}{\overline{D}_{i}}

\newcommand{\Yit}{Y_{it}}
\newcommand{\yit}{y_{it}}

\newcommand{\byi}{\overline{Y}_{i}}

\newcommand{\wit}{W_{it}}

\newcommand{\uit}{U_{it}}

\newcommand{\vit}{V_{it}}

\newcommand{\bvi}{\overline{V}_i}

\newcommand{\bmeta}{\bm{\eta}}

\newcommand{\bmpsi}{\bm{\psi}}

\newcommand{\tl}{\widetilde{l}}

\newcommand{\gi}{\mathbf{g}}

\newcommand{\tm}{\widetilde{m}}

\newcommand{\hmi}{\widehat{m}}
\newcommand{\tmi}{\widetilde{m}}
\newcommand{\bmi}{\overline{m}}

\newcommand{\arginf}{\mathrm{arginf}}

\newcommand{\inv}{^{-1}}

\newcommand{\E}{\mathbb{E}}
\newcommand{\R}{\mathbb{R}}  

\newcommand{\indep}[3]{#1 \perp\kern-5pt \perp #2 \mid #3}

\newcommand{\one}{\mathds{1}}
\newcommand{\I}{\mathbf{I}}

\newcommand{\zero}{\bm{0}}

\newcommand{\Wk}{\mathcal{W}_k}

%% file: 01_introduction.tex
\section{Introduction}
Recent advances in the econometric literature on Machine Learning (ML) use the power of ML algorithms, widely used in data science for solving prediction problems, to enhance existing estimation procedures for treatment and other kinds of causal effect. Notable developments include novel ML algorithms for causal analysis such as Causal Trees by \citep{atheyimbens2016}, Causal Forests by \citep{atheywager2018} and Generalised Random Forests by \citep{atheytibshirani2019}. However, the key development, as far as this paper is concerned, is Double/Debiased Machine Learning (DML) by \citet{chernozhukov2018} wherein ML is used to learn \emph{nuisance functions} with \emph{ex ante} unknown functional forms, and the predicted values of these functions used to construct (orthogonalized) scores for the interest parameters from which consistent and asymptotically normal estimators can be obtained. DML is a very general estimation framework but there are limited examples of its application to panel data, notable examples of which include \citet{chang2020}, \citet{klosin2022}, and \citet{semenova2023}.

In this paper, we develop and assess novel DML procedures for estimating treatment (or causal) effects from panel data with \emph{fixed effects}. The procedures we propose are extensions of the correlated random effects (CRE), within-group (WG) and first-difference (FD) estimators commonly used for linear models to scenarios where the underlying model is non-linear. Specifically, these are based on an extension of the partially linear regression (PLR) model proposed by \citet{robinson1988} to panel data through the inclusion of time-varying predictors and unobserved individual heterogeneity (i.e.\ individual fixed effects). 

Our methodological contribution is twofold and complementary to the recent work on causal panel data estimation using DML by  \citet{chang2020} on difference-in-differences,  \citet{klosin2022}  and  \citep{semenova2023} on high-dimensional treatment heterogeneity in model with fixed effects.

A first contribution is that the procedures we develop are not based on {\em ex ante} taking the nuisance functions or fixed effects to be accurately approximated by high-dimensional sparse functions. Thus, we do not focus solely on the Least Absolute Shrinkage and Selection Operator (LASSO) in the first stage of DML. 
This is in contrast to \citet{klosin2022} and  \citet{semenova2023} who explicitly exploit sparsity to propose alternative procedures.
While we do not rule out using LASSO (and indeed do so in both our simulation study and empirical application), our approach requires only minimal assumptions (like Lipschitz continuity) about the nuisance functions. This is in line with \citet{chang2020}  to encourage researchers to adopt estimation strategies which involve the use of different ML algorithms. Such strategies could involve the use of {\em ensemble} learning such as the stacking of \citet{breiman1996}, the super-learning (a weighted average the best predictions over different choices of ML algorithm) of \cite{laan2007}, or selecting the best-performing learner on an application-by-application basis to ensure inference is based on the most accurate predictions. This is important because, in practice, many ML algorithms are difficult to tune and have been shown to perform very differently for different datasets in different contexts. 

Second, on a computational note, we use \emph{block-k-fold} cross-fitting, where the entire time series of the sampled unit is allocated to one fold to allow for possible serial correlation within each unit as is common with panel data. In this, we are aligned with the concurrent work by \citet{klosin2022} in not needing to rely on the \emph{weak} dependence assumption to deal with dependent data as do \citet{semenova2023}. Conversely, \citet{chang2020} does not require any additional assumption or change in the splitting algorithms because they use only one observation per unit after treatment, ruling out the presence of any fixed effects.

Our method for panel data models with individual fixed effects is general and particularly relevant for applied researchers. We provide new estimation tools within the existing DML framework for use on panel data. In doing so, we broaden the reach of DML to a large family of empirical problems for which the time dimension must be properly accounted for. Our focus, in the subsequent development, on the homogeneous treatment effect case is also because such models are widely used by applied researchers, but we also show how our procedures can be extended to heterogeneous treatment effects provided that the analyst is prepared to specify a (finite-dimensional) parametric model for this heterogeneity. 
More widely, we encourage researchers to use the procedures we propose in place of existing ones, or to test the robustness of their results - based on, say, linear models - to non-linearity.
 
We carry out a simulation study to assess our DML procedures and find large accuracy gains from using DML with flexible learners when the data generating process is highly non-linear (specifically, one involving a non-linear discontinuous function of the regressors); and, while ordinary least squares (OLS) {\em can} outperform ML when the data generating process is linear (as expected) and for some non-linear processes (specifically, for a smooth process that excludes interactions), it is not robust because the analyst never knows the form of the data generating process \emph{ex ante}. Finally, we apply our new procedures to observational panel data by re-analyzing part of the study by \mbox{\citet{fazio2023}} on the effect of the introduction National Minimum Wage (NMW) in the United Kingdom (UK) on voting for conservative parties. We find evidence that the first-difference procedures are the most robust and perform well, and confirm the conclusion from our simulation study that ensemble learning strategies are crucial to obtaining robust results. 

The remainder of the paper is structured as follows. Section~\ref{sec:literature} provides an overview of the literature and places our novel contribution within it. Section~\ref{sec:model} motivates the partially linear panel regression model and the causal assumptions which must hold to ensure the target parameter can be interpreted as a causal effect. Section~\ref{sec:estimators} introduces the two approaches we take to handle the fixed-effects problem.  Section~\ref{sec:estimation} formally introduces the DML estimation procedures. Section~\ref{sec:mcsimul} briefly discusses the Monte Carlo simulation results. Finally, Section~\ref{sec:application} illustrates an empirical application of the procedure and we make concluding remarks in Section~\ref{sec:conclusion}.

\section{Related Literature}\label{sec:literature}
There is a growing body of econometrics literature on using ML for causal inference. One strand focuses on building or modifying existing learners to consistently estimate and make inferences about causal effects \citep[e.g.,][]{atheyimbens2016,atheywager2018,atheytibshirani2019,lechner2022,difrancesco2022}. Another strand focuses on incorporating ML into traditional statistical estimators (e.g., least squares, generalised method of moments, maximum likelihood) to estimate causal effects more accurately \citep[e.g.,][]{belloni2014,belloni2016,chernozhukov2018,chang2020,chernozhukov2022,huber2023}. This paper falls into the second strand.  

Much of the ML literature in econometrics, including that for causal estimation, is built around penalized regression methods like LASSO, where the onus is on the analyst to specify a sufficiently rich data dictionary for the problem at hand. For example, \citet{belloni2016} proposed two-step post-cluster LASSO procedures for panel data with additive individual-specific heterogeneity in which potential control variables are selected using LASSO at stage one followed by the estimation of the homogeneous treatment effect at stage two. 

Two recent papers propose estimation procedures based on LASSO for non-linear panel models. Both focus on different causal targets to ours:\ in the context of continuous treatments, \cite{klosin2022} proposed an estimator for the average partial effect based on a PLR model suitable for static panels, where a first-difference data transformation is used to handle omitted fixed time-invariant confounding; and \citet{semenova2023} propose a procedure for CATE estimation based on a PLR model for dynamic panels. Both of these procedures rely crucially on the key nuisance functions being well approximated by high-dimensional sparse functions:\ without assuming this {\em ex-ante}, their procedures do not follow.
We propose alternative procedures that make no such {\em ex-ante} assumptions, and set these up within the DML framework developed by \citet{chernozhukov2018} who considered the PLR model at length but did not provide any panel examples.

The generality of our approach allows us to 
move beyond LASSO to other types of ML algorithm.  LASSO is powerful but it requires the analyst to pre-specify a sufficiently rich data dictionary, and the memory required to store these in panels with many waves could be prohibitively large such that alternative learners are preferable. However, by staying general, we rule out solutions to the fixed-effects problem like that proposed by \mbox{\citet{semenova2023}} who incorporate the fixed effects into the LASSO data dictionary under a weak sparsity assumption. While recent work by \citet{kolesar2023} argues that, in the absence of prior knowledge to the contrary, this assumption is unlikely to hold, \citet{semenova2023} avoid this issue by decomposing the overall fixed effect into a fixed part, involving the individual-specific means of the (fixed) predictors, and a fixed-effect residual which {\em can} be incorporated into the dictionary under a weak sparsity assumption. 

By not formulating the learning problem in terms of LASSO, we must consider alternative approaches so, instead, propose three estimation procedures that obviate entirely the requirement to model the fixed effects. The first two are based on transforming the outcome to induce a reduced-form model that does not depend on the fixed effects. Because we are agnostic, 
our first-difference procedure differs from that proposed by \citet{klosin2022} in ways we explain further on.  We also consider a procedure based on a correlated random effects model similar to that used by \citet{wooldridge2020} for limited dependent variable panel models to convert the fixed effects into random effects. 

Finally, we note that our work also fits into the literature that leverages the power of ML for causal analysis and policy evaluation. The value added of doubly-robust procedures has been explored in applied works by, for example, \citet{knaus2022double}, \citet{bach2023}, \citet{strittmatter2023}, \citet{baiardi2024plough,baiardi2024}. Because panel data are widely used in applied analyses, our proposed procedures for panel data models have the potential to attract the interest of applied researchers from various fields broadening the applicability of DML. 

%% file: 02_model_estimators.tex
\section{The Partially Linear Panel Regression Model}\label{sec:model}
\setcounter{equation}{0}
\setcounter{theorem}{0}
\setcounter{lemma}{0}
\setcounter{proposition}{0}
\subsection{Notation}
Suppose the panel study collected information on each of $N$ individuals at each of the $t$ time periods, or waves. 
Let $\{(\Yit,\Dit,\vector X_{it}) : \ t=1,\ldots,T\}_{i=1}^{N}$ be $N$ independent and identically distributed (\emph{iid}) random vectors for individual $i$ across all $T$ waves, where $\Yit \in \mathcal{Y}$ is the outcome variable, $\Dit\in{\cal D}$ a continuous or binary treatment  variable (or intervention), and $\vector X_{it} = (X_{it,1},\dots,X_{it,p})'\in \mathcal{X}$ a $p\times 1$ vector of control (pre-determined) variables, usually including a constant term, able to capture time-varying confounding induced by non-random treatment selection.\footnote{Throughout, we use ${\bf v}^\prime$ to indicate the matrix transpose of arbitrary vector {\bf v} and take vectors to be column vectors unless explicitly stated to the contrary.} We denote the realizations of these random variables by $\{(\yit, \dit, \vector x_{it})\}$, respectively. For continuous $\Dit \in \mathcal{D}\subset\R$, if $\dit\geq 0$ then a dose-response relationship is presumed to hold with $\dit=0$ indicating null treatment; otherwise, $\Dit$ is taken to be centered around its mean $\mu_D$ such that $\Dit \equiv \Dit-\mu_D$. For binary $\Dit\in\{0,1\}$, $\dit=0$ is taken to indicate the absence and $\dit=1$ the presence of treatment.

The non-linear model we propose is a simple extension of the partially linear model of \citet{robinson1988} to panel data. We then use a potential outcomes causal framework (e.g.\ \citet{rubin1974}) to set out the assumptions under which its parameters can be interpreted causally. As pointed out by \citet{lechner2015}, this allows us to interpret the resulting partially linear panel model as reduced-form without relying on specific parametric model for the data generating process. To this end, further define the set $\Yit(.)=\{Y_{it}(d):d\in {\cal D}\}$ containing all potential outcomes for individual $i$ at wave $t$, where $\Yit(d)$ is the realization of the outcome for individual $i$ at wave $t$ were the treatment level set to $d$, with one potential outcome for every possible value the treatment could take. The realizations of the wave $t$ potential outcomes are taken to occur before treatment selection at wave $t$, and are linked to the observed outcome by the {\em consistency assumption} that $Y_{it}(\dit)=Y_{it}$ with the others latent {\em counterfactuals}.\footnote{The stable unit treatment value (SUTVA) assumption, that $Y_{it}(d)$ does not depend on the treatment assignments of any other individual, is implicitly taken to hold.} In the interval preceding wave $t$, it is also presumed that the realization of time-varying predictor $\vector X_{it}$ precedes that of $(Y_{it},\Dit)$. 

Finally, define the sets $\xi_i$ of time-invariant heterogeneity terms influencing $(\Yit,\Dit)$, and $L_{t-1}(W_i) = \{W_{i1},\dots,W_{it-1}\}$ the lags of a generic random variable $W_{it}$ available at wave $t$ such that $L_0(W_i)\equiv\varnothing$. 

\subsection{Model and Assumptions}


We extend the partially linear model proposed by \cite{robinson1988} to panel data by introducing fixed effect $\alpha^*_i$ to give the partially linear panel regression  (PLPR) model
\begin{equation}\label{eqn:plriv}
\Yit = \Dit\theta_0  + g_1(\vector X_{it}) + \alpha^*_i + \uit,
\end{equation}
where $g_1$ is a non-linear nuisance function of $\vector X_{it}$ and $E[\uit|\Dit,\vector X_{it},\alpha^*_i]=0$ but $E[\alpha_i^*\mid \Dit,\vector X_{it}]\neq 0$.  The target parameter $\theta_0$ is the average partial effect of continuous $\Dit$ such that $d\theta_0=E[\Yit(d)-\Yit(0)]$, or the average treatment effect (ATE) $E[\Yit(1)-\Yit(0)]$ for binary treatments.

Following \cite{chernozhukov2018}, we focus on the partialled-out PLPR (PO-PLPR) model
\begin{align}
\Yit &= \vit\theta_0  + l_1(\vector X_{it}) + \alpha_i +  \uit\label{eqn:plr_y},\\
\vit &= \Dit - m_1(\vector X_{it}) -\gamma_i,\label{eqn:plr_v}
\end{align}
where $l_1$ and $m_1$ are nuisance functions, $\alpha_i$ is a fixed effect, $E[\uit\mid\vit,\vector X_{it},\alpha_i]=0$, and $\vit$ the residual of a non-linear additive noise treatment model depending on fixed effect $\gamma_i$ and satisfying $E[\vit|\vector X_{it},\gamma_i]=0$. The focus will be on the PO-PLPR 
rather than the PLPR model 
because, as shown below, $l_1$ is a conditional expectation over $\Yit$ whereas $g_1$ is a conditional expectation over counterfactual $\Yit(0)$ that must generally be learnt iteratively. Moreover, the orthogonalized score on which DML estimation of $\theta_0$ is based involves $\vit$ whether it is derived under PLPR or PO-PLPR (see Section \ref{sec:estimation}).


We now detail the underlying assumptions required for $\theta_0$ to be feasibly estimable with a causal interpretation. Note that some of these assumptions are stated in terms of stochastic conditional independence ${\perp\!\!\!\perp}$ to simplify the development. Recalling that $\xi_i$ represents the omitted time-invariant confounding variables, the first two assumptions can be stated as follows:
\begin{assumption}\label{item:asm_feedback} 
    \textsc{(No feedback to predictors)} $\vector X_{it} \indep{}  {L_{t-1}(Y_i,D_i)}   L_{t-1}(\vector X_i),\xi_i.$ 
\end{assumption}
\begin{assumption}\label{item:asm_nolags} 
    \textsc{(Static panel)} $\Yit,\Dit  \indep{}  {L_{t-1}(Y_i,\vector X_i,D_i)} \vector X_{it},\xi_i$. 
\end{assumption}
\noindent Assumptions~\ref{item:asm_feedback}-\ref{item:asm_nolags} together ensure that the panel is static and that any observed lag dependence is due to non-causal autocorrelation. Specifically, these assumptions ensure that the joint distribution of outcomes and treatments given the time-varying predictors and omitted time-invariant influences satisfies 
\begin{equation*}
p\{Y_{i1},D_{i1},\dots,Y_{iT},D_{iT}|L_T(\vector X_i),\xi_i\}=\prod_{t=1}^Tp(\Yit,D_{it}|\vector X_{it},\xi_i),
\end{equation*}
where $p(.)$ denotes a density function for the (conditional or joint) distribution indicated by its arguments. Moreover, there is no need to model the distribution of the time-varying predictors, and the {\em initial conditions problem}, which would arise were the panel study to have started after the joint process began, is avoided.  

To give $\theta_0$ a causal interpretation, we require that the data generating process for the potential outcomes and treatment satisfies the following conditions:
\begin{assumption}\label{item:asm_selection_fe} 
    \textsc{(Selection on observables and omitted time-invariant variables)}
    $\Yit(.)\indep{}\Dit \vector X_{it},\xi_i$.    
\end{assumption}
\begin{assumption} \label{item:asm_effect_fe}   
    \textsc{(Homogeneity and linearity of the treatment effect)}  \\
    $E[\Yit(d)-\Yit(0)|\vector X_{it},\xi_i]=d\theta_0$.
\end{assumption}
Assumption~\ref{item:asm_selection_fe} states that treatment selection at wave $t$ is (strongly) ignorable given $\vector X_{it}$ and latent $\xi_i$.  Assumption~\ref{item:asm_effect_fe} requires that 
$Y_{it}(d)-Y_{it}(0)$ is constant or varies between individuals independently of $\vector X_{it}$ and $\xi_i$. Alternatively, under the weakly ignorable assumption that only $\Yit(0)\indep{}\Dit \vector X_{it},\xi_i$, target parameter $d\theta_0=E[\Yit(d)-\Yit(0)\mid \Dit=d]$.  However, we assume throughout that the strong version holds. 

The final assumption is that the combined effect of the confounding variables is additively separable into that of the observed confounding variables and that of the omitted time-invariant confounding variables.

\begin{assumption}\label{item:asm_additive} 
    \textsc{(Additive Separability)} 
    \begin{enumerate}[label=(\alph*)]
        \item $E[\Yit(0)\mid\vector X_{it},\xi_i]=g_1(\vector X_{it})+\alpha^*_i$
        where $\alpha^*_i=\alpha^*(\xi_i),$ 
        \label{item:plpr}
        \item $E[\Dit\mid\vector X_{it},\xi_i]=m_1(\vector X_{it})+\gamma_i$
        where $\gamma_i=\gamma(\xi_i)$.\label{item:po-plpr}
    \end{enumerate}
\end{assumption}
\noindent Assumptions~\ref{item:asm_selection_fe} and \ref{item:asm_effect_fe} imply that $E[\Yit\mid \Dit,\vector X_{it}, \xi_i]=E[\Yit(0)\mid\vector X_{it},\xi_i]+\Dit\theta_0$. Hence, Assumption~\ref{item:asm_additive}\ref{item:plpr} leads instantly to PLPR model \eqref{eqn:plriv}. Moreover, Assumptions~\ref{item:asm_selection_fe} and~\ref{item:asm_effect_fe} also imply that $E[\Yit(0)\mid \vector X_{it}, \xi_i]= E[\Yit\mid\vector X_{it},\xi_i]- E[\Dit\mid\vector X_{it},\xi]\theta_0$, from which it follows from both Assumptions~\ref{item:asm_additive}\ref{item:plpr} and~\ref{item:asm_additive}\ref{item:po-plpr} that PO-PLPR model (\ref{eqn:plr_y})-(\ref{eqn:plr_v}) holds because $E[\Yit\mid\vector X_{it},\xi_i]=l_1(\vector X_{it})+\alpha_i$, where $l_1(\vector X_{it})=g_1(\vector X_{it})+m_1(\vector X_{it})\theta_0$ and $\alpha_i=\alpha^*_i+\gamma_i\theta_0$. \\

In the description of each approach, below, we focus on procedures for the homogenous treatment effects case. However, all of these procedures can be extended to heterogeneous treatment effects provided the analyst is prepared to specify a (finite-dimensional) parametric heterogeneity model. A summary of how this is done can be found in Section~\ref{sec:hetero}. 

\section{Fixed-effects Estimation}\label{sec:estimators}
\setcounter{equation}{0}
\setcounter{theorem}{0}
\setcounter{lemma}{0}
\setcounter{proposition}{0}

We propose two approaches for estimating $\theta_0$ from PO-PLPR model \eqref{eqn:plr_y}-\eqref{eqn:plr_v}. The presence of the time-invariant fixed effects means that only $l_1(\vector X_{it})+E[\alpha_i|\vector X_{it}]$ and $m_1(\vector X_{it})+E[\gamma_i|\vector X_{it}]$ can be learnt from the observed data directly through $E[\Yit\mid\vector X_{it}]$ and $E[\Dit\mid\vector X_{it}]$, respectively, so the first challenge is to remove the fixed effects just as it is in the linear case. As such, the approaches we propose are adaptations of existing techniques for {\em linear} panel data models. The first is based on correlated random effects (CRE) and the second on two varieties of data transformation:\ first-difference (FD) and within-group (WG).\footnote{Under the \emph{random effects assumption}, where $E[\alpha_i\mid\vector X_{it}]=E[\alpha_i]=0$ and $E[\gamma_i\mid\vector X_{it}]=E[\gamma_i]=0$, $l_1$ and $m_1$ can be straightforwardly learnt from the observed data, but we assert that this is unlikely to hold in practice and, hence, causal inference would not be credible. However, were the analyst prepared to make the random effects assumption, we note that \citet{sela2012} developed an algorithm for using tree-based learners to estimate non-causal partially linear regression models.}  

Crucially, to understand our contribution, it is important to note that the use of CRE or data transformations does not in itself lead to a consistent estimator of $\theta_0$. There are further challenges, created by the non-linearity of the nuisance functions $l_1$ and $m_1$, to be overcome. Specifically, we do not solve these challanges by {\em a priori} taking $l_1$ and $m_1$ to be accurately approximated by high-dimensional sparse functions of the form $l_1(\vector X_{it})\approx \vector B^{\prime}_{it}\pmb{\psi}$ and $m_1(\vector X_{it})\approx \vector B^{\prime}_{it}\pmb{\phi}$, where vector ${\bf B}_{it}={\bf B}(\vector X_{it})$ is an analyst-specified high-dimensional data dictionary of linear and non-linear terms, $\pmb{\psi}$ and $\pmb{\phi}$ are vectors of parameters with ${\rm dim}(\pmb{\psi})>>||\pmb{\psi}||_0$ and ${\rm dim}(\pmb{\phi})>>||\pmb{\phi}||_0$, and $||{\bf v}||_0$ is the number of non-zero elements in arbitrary vector ${\bf v}$. The FD and CRE procedures respectively developed by \cite{klosin2022} and \cite{semenova2023} rely explicitly on such representations. However, our procedures are suitable for DML based on ensemble learning strategies. 


\subsection{Correlated Random Effects}\label{sec:three_approaches}

Correlated random effects (CRE) models are extensions of fixed effect models in which the fixed effect is replaced by a model for its dependence on the predictor variables. CRE models can depend on any function of $\{\vector X_{it}:t=1,\dots,T\}$ (e.g.\ \cite{wooldridge2020}) but practice tends to focus on {\em exchangeable} functions and especially functions of $\overline{\vector X}_i=T^{-1}\sum_{t=1}^T\vector X_{it}$ following \cite{mundlak1978} who first demonstrated the equivalence of CRE and fixed effects estimators when the effects of $\vector X_{it}$ in the fixed-effects model and the effects of ${\overline {\vector X}}_i$ in the model for the fixed effect are all linear.

In general, the fixed-effect model must be correctly specified, but a major advantage of the Mundlak device for linear panel models, where the effects of $\vector X_{it}$ are linear, is that least-squares estimators of the coefficients remain consistent even if the linear model for $E[\alpha_i|{\overline {\vector X}}_i]$ is mis-specfied, provided that $\alpha_i$ and ${\overline {\vector X}}_i$ are correlated. However, this robustness is lost for PO-PLPR model (\ref{eqn:plr_y})-(\ref{eqn:plr_v}) because $l_1$ and $m_1$ are non-linear. This is also the case for the linear panel model with random coefficients for heterogeneous treatment effects of \cite{wooldridge2019} and the non-linear probit model of \cite{wooldridge2020}.


In the spirit of \cite{wooldridge2020}, we derive a CRE model based on PO-PLPR model (\ref{eqn:plr_y})-(\ref{eqn:plr_v}). Suppose that the data generating process 
satisfies Assumptions~\ref{item:asm_feedback}-\ref{item:asm_additive} 
and the fixed effects follow non-linear additive noise models $\alpha_i= \omega_\alpha({\overline{\vector X}_i})+a_i$  and  $\gamma_i = \omega_\gamma({\overline{\vector X}_i}) + c_i$, where $a_i$ and $c_i$ are random effects satisfying $(a_i,c_i)\ {\perp\!\!\!\perp}\ L_T(\vector X_i)$ and $a_i\ {\perp\!\!\!\perp}\ c_i$. Then the CRE model with nuisance parameters $\widetilde{l}_1$ and $\widetilde{m}_1$ is
\begin{align} 
    \Yit & =\vit\theta_0 + \widetilde{l}_1(\vector X_{it},{\overline {\vector X}}_i)+a_i+\uit \label{eqn:cre_y}\\ 
    \vit & = \Dit - \widetilde{m}_1(\vector X_{it},{\overline{\vector X}_i}) - c_i, \label{eqn:cre_v} 
\end{align} 
\noindent where $E[\uit \mid\vit,\vector X_{it},{\overline{\vector X}_i},a_i\big]=E[\vit|\vector X_{it},{\overline{\vector X}_i},c_i]=0$, $\widetilde{l}_1(\vector X_{it},{\overline{\vector X}_i})=l_1(\vector X_{it}) + \omega_{\alpha}({\overline{\vector X}_i})$ and $\widetilde{m}_1(\vector X_{it},{\overline{\vector X}_i}) = m_1(\vector X_{it})+\omega_\gamma({\overline{\vector X}_i})$. The random effects simply capture autocorrelation between observations on the same individual. Crucially, $\widetilde{l}_1$ and $\widetilde{m}_1$ can be learnt directly from $E[\Yit\mid\vector X_{it}, {\overline{\vector X}_i}]$ and $E[\Dit\mid\vector X_{it}, {\overline{\vector X}_i}]$, respectively.\footnote{\cite{semenova2023} also use a CRE model in their formulation for dynamic models.  In our static panel case with only {\em fixed} $\vector X_{it}$ predictors, this would be equivalent to specifying $l_1(X_{it})+\alpha_i=\psi_0 \overline{X}_i + \alpha_i$ with $(\alpha_1,\dots,\alpha_N)$ assumed to be a {\em weakly sparse} vector following a Mundlak-style model which can be estimated using LASSO. Alternative models for weakly sparse fixed effects have been developed by \cite{kock2016} and \cite{kock2019} which can be estimated using {\em de-sparsified} LASSO.} 

\subsection{Data Transformations}\label{sec:exact}
The second approach we consider follows more conventional techniques for panel data by transforming the data to remove entirely the fixed effects from the analysis. 

Let $\wit$ be a generic random variable and $Q$ a panel data transformation operator such that $Q(\wit)=Q_t(W_{i1},\dots,W_{iT})$ is a function of the random variable at wave $t$ and the remaining realizations for \mbox{individual $i$}. We consider two such transformations:\ the WG (within-group) or time-demeaning transformation, i.e., $Q(\wit)=\wit-\overline{W}_i$, where  $\overline{W}_{i}= T\inv \sum_{t=1}^T \wit$; and the FD (first-difference) transformation, i.e., $Q(\wit) = \wit-W_{it-1}$ for $t=2,\dots,T$. 

The reduced-form model for $Q(\Yit)$ and $Q(\vit)$ under PO-PLPR model~\eqref{eqn:plr_y}-\eqref{eqn:plr_v} is 
\begin{align}
    Q(\Yit) & = Q(\vit)\theta_0 + Q\big({l}_1(\vector X_{it})\big) + Q(\uit) \label{eqn:Q_y} \\
    Q(\vit) & = Q(\Dit)-Q\big({m}_1(\vector X_{it})\big), \label{eqn:Q_d}
\end{align}
\noindent which does not depend on fixed effects $\alpha_i$ and $\gamma_i$ because $Q(\alpha_i)=Q(\gamma_i)=0$.

The challenge of learning the transformed nuisance functions $Q({l}_1(\vector X_{it}))$ and $Q({m}_1(\vector X_{it}))$ is complicated by the non-linearity of $l_1$ and $m_1$ because, as set out previously, we do not {\em a priori} assume that the nuisance functions admit high-dimensional sparse representations of $l_1$ and $m_1$ where $Q(l_1(\vector X_{it})\approx Q({\bf B}^{\prime}_{it})\pmb{\psi}$ and $Q(m_1(\vector X_{it}))\approx Q({\bf B}^{\prime}_{it})\pmb{\phi}$ could be learnt directly as per the FD procedure proposed by \cite{klosin2022}. 


\subsection{Heterogeneity}\label{sec:hetero}


If Assumption~\ref{item:asm_effect_fe} is implausible or the inferential target of the analysis is heterogeneity of the treatment effects itself then further modelling is required. Completely relaxing Assumption~\ref{item:asm_effect_fe} results in $\theta_0d=E[\Yit(d)-\Yit(0)\mid\vector X_{it},\xi_i]$, that is, conditional average partial effects or conditional average treatment effects (CATE) dependent on all of the confounding variables. Focussing on the binary treatment case for simplicity, extending Assumption \ref{item:asm_additive}\ref{item:plpr} so that $E[\Yit(1)\mid\vector X_{it},\xi_i]$ is also additively separable would lead to an additively separable CATE $\theta_0=h_0(\vector X_{it})+\delta_i$, where $\delta_i=\delta(\xi_i)$ is what \cite{wooldridge2019} refers to as a {\em random coefficient}.

CRE model (\ref{eqn:cre_y})-(\ref{eqn:cre_v}) can be specified in terms of CATEs if Assumption~\ref{item:asm_effect_fe} fails. Temporarily discarding vector notation, if $E[\Yit(1)-\Yit(0)\mid X_{it},\xi_i]=h_0(X_{it})$ then our procedure allows parametric model $h_0(X_{it})=S_{it}\theta_0$, where $S_{it}=S(X_{it})$ captures treatment effect heterogeneity though interaction $S_{it}\vit$ replacing $\vit$ in (\ref{eqn:cre_y}). If the heterogeneity also depends on $\xi_i$ then, following \cite{wooldridge2019}, the analyst can proceed by additionally specifying a further linear model for $E[\delta_i\mid \overline{X}_i]$ so that $S_{it}$ is extended to $S_{it}=S( X_{it},\overline{X}_i)$ with no interactions between $X_{it}$ and $\overline{X}_i$ included. As noted by \cite{wooldridge2019}, this model must be correctly specified to induce a consistent estimator, despite having the appearance of the Mundlak device.

Finally, for the transformation models, the failure of Assumption~\ref{item:asm_effect_fe} also requires the analyst to specify parametric model $h_0(X_{it})=S_{it}\theta_0$. Then \eqref{eqn:Q_y} becomes $Q(\Yit) = Q(\vit S_{it})\theta_0 + Q\big({l}_1(\vector X_{it})\big) + Q(\uit)$. In contrast to the CRE, where the analyst must assume $\delta_i=0$ or specify the model $E[\delta_i\mid{\overline{\vector X}}_i]$, random coefficient $\delta_i$ cancels out.

Our approach contrasts with those of \cite{klosin2022} and \cite{semenova2023}, both of which are based on learning the high-dimensional sparse representation of $h_1$ using a suitably specified data dictionary.  The former is a FD procedure that removes $\delta_i$ from consideration provided the CATE is additively separable as above, whereas the latter implicitly assumes that $\delta_i=0$.

%% file: 03a_estimation_inference.tex
\section{DML procedures for Panel Data}\label{sec:estimation}
\setcounter{equation}{0}

We now describe our DML procedures for the fixed-effects estimators presented in Section~\ref{sec:estimators}. 
The objective is to make inferences on the target parameter $\theta_0$ given a suitable predictions of nuisance functions $\bmeta_0$ based on the observed data $W_i = \{W_{it}: t=1,\dots,T\}$, where $W_{it} = \{\Yit,\Dit,\vector X_{it}, \overline{\vector X}_i\}$ 
for CRE, and $W_{it} = \{Q(\Yit),Q(\Dit),Q(\vector X_{it})\}$ for the transformation approach (noting that $W_{i1}\equiv\varnothing$ if $Q$ is the FD transformation). 

\subsection{Learning the Nuisance Parameters}\label{sec:nuisance}
\setcounter{equation}{0}
\setcounter{theorem}{0}
\setcounter{lemma}{0}
\setcounter{proposition}{0}
The first stage of DML involves using a suitably chosen learner (or combination of learners) to predict flexibly the unknown nuisance parameters. The nuisance functions $\bmeta_{0i}=\bmeta_0(\vector X_{i1},\ldots,\vector X_{iT})$ vary between individuals and so the task is to learn $\bmeta_0$.  As discussed previously, the presence of the fixed effects and non-linear functional forms pose challenges which need to be addressed. Below we describe different procedures for learning $\bmeta_0$. 

\subsubsection{Correlated Random Effects}\label{sec:learning_cre}  

The estimation of $\theta_0$ based on model (\ref{eqn:cre_y})-(\ref{eqn:cre_v}) requires learning $\widetilde{l}_1(\vector X_{it},\overline{\vector X}_i)$ from the data $\{\Yit,\vector X_{it},\overline{\vector X}_i:t=1,\dots,T\}_{i=1}^N$, and obtaining predicted residual $\widehat{V}_{it}$ to plug into Equation~\eqref{eqn:cre_y}. Simply learning $\widetilde{m}_1(\vector X_{it},\overline{\vector X}_i)$ from the data $\{\Dit,\vector X_{it},\overline{\vector X}_i:t=1,\dots,T\}_{i=1}^N$ and using $\widehat{V}_{it}=\Dit - {\widehat{m}_1}(\vector X_{it},\overline{\vector X}_i)$ would ignore individual random effect $c_i$ and so introduce bias. Hence, we propose the following approach:
\setcounter{bean}{0} 
\begin{center} 
\begin{list} 
{\textsc{Step} \arabic{bean}.}{\usecounter{bean}} 
\item Learn $\tmi_1(.)$ from $\big\{\Dit,\vector X_{it},\overline{\vector X}_i:t=1,\dots,T \big\}_{i=1}^N$ with prediction \\$\widehat{m}_{1i}=\tmi_1\widehat{(\vector X_{it},\overline{\vector X}_i)}$. 
    \item Calculate $\widehat{\bmi}_i=T\inv\sum_{t=1}^T\hmi_{1i}$. 
    \item Calculate $\widehat{m^*_1}{(\vector X_{it},\overline{\vector X}_i,\bdi)}=\widehat{m}_{1i}+\bdi-\widehat{\bmi}_i$,
\end{list} 
\end{center} 
\noindent where $m^*_1(\vector X_{it},\overline{\vector X}_i,\bdi)=E[\Dit\mid \vector X_{it},\overline{\vector X}_i]+c_i$. 
To capture $c_i$, it is essential to include the individual-specific treatment mean $\bdi$ to ensure $\vit$ is included in the prediction of $m_1^*$. For example, if the conditional distribution $D_{i1},\dots,D_{iT}\mid\vector X_{i1},...,\vector X_{iT}$ is multivariate normal with $E[\Dit|\vector X_{it},\overline{\vector X}_i] = \tmi_1(\vector X_{it},\overline{\vector X}_i)$ and a homoskedastic variance-covariance matrix with random effects structure (induced by marginalizing over $c_i$) then $E(\Dit\mid\vector{X}_{it},\overline{\vector X}_i,\overline{D}_i)=\widetilde{m}_1(\vector X_{it},\overline{\vector X}_i)+c_i=m_1^*$ as required, but it is generally necessary to follow the steps above. More details are given in Section~\ref{sec:proof_lemma_cre} of the Online Supplementary Information.  


\

\subsubsection{Transformation approaches}\label{sec:learning_transf}   

First, we consider learning based on the transformed data alone. Generally, $Q\big({l}_1(\vector X_{it})\big)\ne {l}_1\big(Q(\vector X_{it})\big)$ and $Q\big({m}_1(\vector X_{it})\big)\ne {m}_1\big(Q(\vector X_{it})\big)$ unless $l_1$ and $m_1$ are linear. This means a consistent estimator cannot straightforwardly be constructed from the transformed data. However, we propose the following procedures: 

\emph{(a) Approximate procedure.} Approximate model (\ref{eqn:Q_y})-(\ref{eqn:Q_d}) by 
\begin{align} 
       Q(\Yit) & \approx Q(\vit)\theta_0 + {l}_1\big(Q(\vector X_{it})\big) + Q(\uit) \label{eqn:aQ_y}\\ 
       Q(\vit) & \approx  Q(\Dit) - {m}_1\big(Q(\vector X_{it})\big), \label{eqn:aQ_v} 
\end{align} 
\noindent where $l_1$ and $m_1$ can be learnt from the transformed data $\{Q(\Yit),Q(\vector X_{it}): t=1,\dots,T\}_{i=1}^N$ and $\{Q(\Dit),Q(\vector X_{it}): t=1,\dots,T\}_{i=1}^N$, respectively. 
\noindent This approach can produce good approximations of some non-linear nuisance functions but it is not robust and we were able to choose functions where it performed poorly. 

\emph{(b) Exact procedure.} Consider first the FD transformation $Q(\Yit)=\Yit-Y_{it-1}$. Then under Assumptions \ref{item:asm_feedback}-\ref{item:asm_additive}, $E[\Yit-Y_{it-1} \mid \vector X_{it-1} , \vector X_{it}]=\Delta l_1(\vector X_{it-1} ,\vector X_{it})$ and  $E[\Dit-D_{it-1}\mid\vector X_{it-1} ,\vector X_{it}]=\Delta m_1(\vector X_{it-1} ,\vector X_{it})$, so that $\Delta l_1(\vector X_{it-1} ,\vector X_{it})$ can be learnt using $\{\Yit-Y_{it-1},\vector X_{it-1} , \vector X_{it}: t=2,\dots,T\}_{i=1}^N$ and $\Delta m_1(\vector X_{it-1} ,\vector X_{it})$ can be learnt using $\{\Dit-D_{it-1}, \vector X_{it-1} , \vector X_{it}:t=2,\ldots,T\}_{i=1}^N$.  However, for WG transformation $E[\Yit-\byi\mid L_T(\vector X_i) ]=l_1(\vector X_{it}) - T\inv\sum_{s=1}^T l_1(\vector X_{is})$ and $E[\Dit-\bdi| L_T(\vector X_i)]=m_1(\vector X_{it}) - T\inv\sum_{s=1}^T m_1(\vector X_{is})$, the dimension of the learning problem is $O(T^2)$ times greater than that of the FD transformation and so unfeasible for non-trivial $T$. 

\emph{(c) Hybrid procedure.}  Suppose the conditions set out in Section~\ref{sec:three_approaches} for model~\eqref{eqn:cre_y}-\eqref{eqn:cre_v} hold. Then $\tl_1(\vector X_{it},\overline{\vector X}_i)=l_1 (\vector X_{it})+\omega_\alpha(\overline{\vector X}_i)$ and $\tm_1(\vector X_{it},\overline{\vector X}_i)=m_1(\vector X_{it})+\omega_\gamma(\overline{\vector X}_i)$ learnt from the sample data satisfy $Q(\tl_1(\vector X_{it},\overline{\vector X}_i))=Q(l_1(\vector X_{it}))$ and $Q(\tm_1(\vector X_{it},\overline{\vector X}_i)\big)=Q(m_1(\vector X_{it})$. The hybrid approach can be used with both WG and FD transformations. 

We elaborate on the justification for these procedures in Sections~\ref{sec:proof_lemma_approx} and~\ref{sec:proof_lemma_exact} of the Online Supplementary Information.

\subsection{Neyman Orthogonal Score Function}\label{sec:no_score}
The second stage of DML requires specifying an orthogonal score function for the structural parameter of interest to be solved after the nuisance functions have been plugged in. Following \cite{chernozhukov2018}, we construct a generic Neyman orthogonal score function for panel data that accounts for (a) the presence of the unobserved individual heterogeneity, and (b) non-linearity of the nuisance functions. Its properties allow us to obtain valid inferences using DML algorithms, provided the ML algorithms converge at rate $N^{1/4}$.
We wish to make inference on the target parameter $\theta_0$ given a suitable estimate of the nuisance parameter $\bmeta_0$ from the data. We repurpose the definition of $W_{it}$ to define the entire dataset as $W = \cup^{N}_{i=1}W_i$, where $W_i=\{W_{it}: t=1,\dots,T\}$, $W_{it} = w\{\Yit,\Dit,\vector X_{it}\}$ and $w$ is a function or transformation of the data (possibly the identity) chosen by the analyst to implement one of the estimation approaches discussed in Section~\ref{sec:estimators}. 


For individual $i$, let ${\bf r}_i=(r_{i1},...,r_{iT})^\prime$ be the $T$ residuals of the PO-PLPR model (either (\ref{eqn:cre_y})-(\ref{eqn:cre_v}) or  (\ref{eqn:aQ_y})-(\ref{eqn:aQ_v})), 
$X_i=X(\vector X_{i1},\dots,\vector X_{iT})$ the appropriately chosen set of
predictor variables, and $\Sigma(X_i)=E[{\bf r}_i{\bf r}'_i\mid X_i]$ the (potentially) heteroskedastic residual variance-covariance matrix. Note that $r_{it}$ differs from $\uit$ in that it potentially incorporates the model random effect and other sources of autocorrelation.

Then the Neyman orthogonal score has the form 
\begin{equation}\label{eqn:sec_score} 
    \psi^\perp(W_i;\theta_0,\bmeta_0)={\bf V}_i^\perp\Sigma_0\inv(X_i){\bf r}_i, 
\end{equation} 
\noindent where row vector ${\bf V}^\perp_i=(V_{i1}^\perp,\ldots, V^\perp_{iT})$ contains the {\em orthogonalized regressors} chosen to ensure Neyman orthogonality.
\begin{enumerate}[label=(\alph*)]
\item For CRE: Under model (\ref{eqn:cre_y})-(\ref{eqn:cre_v}), $r_{it}=a_i+U_{it}=\Yit-\vit \theta_0 - \widetilde{l}_1(\vector X_{it},\overline{\vector X}_{i})$, $V_{it}=\Dit - \widetilde{m}_1(\vector X_{it},\overline{\vector X}_i) -c_i$ and $V_{it}^\perp=V_{it}$.  
\item For FD: Under model (\ref{eqn:Q_y})-(\ref{eqn:Q_d}),  $r_{it}=Q(u_{it})=Q(\Yit)-Q\big(\vit\big)\theta_0 - Q\big(l_1(\vector X_{it}) \big)$ and $V_{it}^{\perp}=Q(V_{it})$.  
\end{enumerate}

A derivation of~\eqref{eqn:sec_score} and its generalization to estimating conditional average treatment effects (CATEs) in the presence of heterogeneous treatment effects is provided in Section~\ref{sec:proof:score} of the Online Supplement.
\subsection{Estimation and Inference About the Target Parameter}\label{sec:estimation_para}

Suppose that 
suitable ML algorithms are available to learn the nuisance functions and obtain estimate ${\widehat\bmeta}$ on a pathway converging to $\bmeta_0$ at rate $N^{1/4}$. We now set out the DML2 \cite[Def.\ 3.2]{chernozhukov2018} algorithm we use to make post-regularized inference about $\theta_0$ based on Neyman-orthogonal score (\ref{eqn:sec_score}).


DML uses cross-fitting to control the overfitting or regularization bias introduced by plugging in ${\widehat\bmeta}$ to (\ref{eqn:sec_score}). A key feature of our procedure is that we have adapted DML to use \emph{block-k-fold} cross-fitting for panel data. This involves treating the time-series for each sample unit, $W_i$, as {\em k-fold}-sampling units when partitioning the sample into $K$ equi-sized folds. Because $W_i$ satisfies the independent and identically distributed assumption, whereas $W_{it}$ does not, block cross-fitting accounts for the within-individual autocorrelation structure and so avoids having to make further post-fitting adjustments for autocorrelation to the estimmated standard errors as did \cite{semenova2023}.


Letting ${\cal W}=\{1,\ldots, N\}$ denote the indices of the sample units, our DML estimation procedure with \emph{block-k-fold} cross-fitting can be set out as follows:

\input{03b_dml_algorithm}


Because score (\ref{eqn:sec_score}) is linear in $\theta$, 
our proposition is that $\widehat{\theta}^{DML}$ converges to normality according to the vectorized equivalent of \citet[Theorem 4.1] {chernozhukov2018} under the regularity conditions set out by \citet[Assumption 4.1]{chernozhukov2018}. 

%% file: 03b_dml_algorithm.tex

\begin{algorithm}[\textsc{DML estimator}] \label{dml_algorithm}
\textnormal{}
\begin{center}
\begin{list}
{\textsc{Step} \arabic{bean}.}{\usecounter{bean}}
    \item  \textnormal{\textsc{(Sample splitting and cross-fitting)} Randomly partition the cross-sectional units in the estimating sample into $K$ folds of the same size. Denote the units in fold \mbox{$k=1,\ldots,K$} by $\Wk\subset\mathcal{W}$ and let $\Wk^c$ be its complement such that $N_k\equiv |\Wk| = N/K$, $|\Wk^c|=N-N_k$ and, because the folds are mutually exclusive and exhaustive, $\Wk\cap\mathcal{W}_j=\Wk\cap\Wk^c=\varnothing$ and $\Wk\cup\Wk^c=\mathcal{W}_1\cup\ldots\cup\mathcal{W}_K=\mathcal{W}$. For \mbox{$K>2$}, the larger complementary sample $\Wk^c$ is used to learn the potentially complex nuisance parameters $\bmeta$, and $\Wk$ for the relatively simple task of estimating the target parameter $\theta_0$. 
    }
    
    \vspace{2mm}
 
    \item  \textnormal{\textsc{(Nuisance learning)} For fold $k$, learn $\bmeta_0$ from the data $\{W_i:i\in \Wk^c\}$ using one of the approaches from Section~\ref{sec:nuisance}. Then use the learnt prediction rule $\widehat{\bmeta}_k$ to construct $\bmpsi^\perp(W_k; \theta, \widehat{\bmeta}_k)$ for each fold.}\\

    \vspace{2mm}

    \item  \textnormal{\textsc{(Estimation)}  The DML estimator $\widehat{\theta}^{DML}$ is obtained by solving 
    \begin{equation}\label{eqn:fmom_cond}
    \frac{1}{K}\sum_{k=1}^K\frac{1}{N_k}\sum_{i\in \Wk} \bmpsi^\perp(W_i; \widehat{\theta}^{DML}, \widehat{\bmeta}_k)  = \zero,
    \end{equation}
    \noindent which has closed-form solution
      \begin{equation}
       \widehat{\theta}^{DML} = \Bigg(\sum_{k=1}^K\sum_{i\in\Wk}\widehat{\vector  V}_i'\widehat{\vector  V}_i \Bigg)\inv %
       \sum_{k=1}^K\sum_{i\in\Wk} \widehat{\vector  V}_i'\widehat{\vector U}_i,
      \end{equation}
    \noindent  where $\widehat{\vector  V}_i = (\widehat{V}_{i1},\dots,\widehat{V}_{iT})'$ and $\widehat{\vector U}_i = (\widehat{U}_{i1},\dots,\widehat{U}_{iT})'$. 
    Note that the closed-form expression is obtained by setting $\Sigma_0 = I_T$ in equation~\eqref{eqn:sec_score} and capturing autocorrelation using cluster-robust versions of the following variance estimators. 
    The estimator of the variance of $\widehat{\theta}^{DML}$ is
    \begin{equation}
        \widehat{\sigma}^2 = \widehat{J}\inv\Bigg\{\frac{1}{K}\sum_{k=1}^K\frac{1}{N_k}\sum_{i\in \Wk}\psi^\perp({W}_i;  \widehat{\theta}^{DML}, \widehat{\bf{\bmeta}}_{k})\psi^\perp(W_i; \widehat{\theta}^{DML}, \widehat{\bmeta}_{k})'\Bigg\} \widehat{J}_k\inv
    \end{equation}
    \noindent where $\widehat{J} =K\inv\sum_{k=1}^KN_k\inv\sum_{i\in \Wk}\widehat{\vector  V}_i'\widehat{\vector  V}_i$.
    \footnote{We additionally make a finite-sample correction $(\widehat{\theta}^{DML}-K\inv\sum_k\thetano_k)^2$ 
    to $\widehat{\sigma}^2$, where $\thetano_k=\big(\sum_{i\in\Wk}\widehat{\vector  V}_i'\widehat{\vector  V}_i\big)\inv \sum_{i\in\Wk} \widehat{\vector  V}_i'\widehat{\vector U}_i$ is the fold-specific estimate as in \citet[p.~C30]{chernozhukov2018}.}    
    }\\

    \vspace{2mm}

    \item \textnormal{\textsc{(Iteration)} Repeat \textsc{Steps}~1-3 for each of the $k$-folds and average out the results.}
\end{list}
\end{center}
\end{algorithm}

%% file: 04_simulations.tex
\section{Simulation Study}\label{sec:mcsimul}
\setcounter{equation}{0}
The primary focus of our Monte Carlo simulation study is to assess the performance of the DML procedures we propose in terms of bias and precision for different ML algorithms.  We generated data under DGPs satisfying the assumptions required for the PO-PLPR model to hold for three different pairs of nuisance functions:\ linear, non-linear (smoothly continuous with no interactions), and non-linear (discontinuous with interactions). These are described along with a detailed discussion of the results in Online Supplement~\ref{sec:appx_mc_simul}. 
The DGP satisfies the usual sparsity constraints because only two of the thirty predictors included in the analysis have non-zero effects. 

Table~\ref{tab:mx_collapse} provides a summary of the main results under the most complex non-linear DGP for four of the procedures introduced above (Panels A-C). For each procedure, standard OLS estimates are compared with DML using four different learners for the non-linear nuisance functions. Across all three procedures, OLS can exhibit large bias and DML-LASSO typically outperforms the other learners in terms of both bias and precision, with the exception of the WG estimator (Panel C). In practice, LASSO requires the analyst to specify a sufficiently rich dictionary of non-linear terms; had only linear terms been included, for example, its performance would have been closer to OLS. Gradient boosting outperforms the other tree-based approaches (CART and RF) the performance of which in terms of bias and precision (SE/SD) is far inferior. A closer investigation, described in Online Supplement~\ref{sec:simresults}, found the sampling distributions of tree-based ${\hat{\theta}}_{DML}$ to be highly non-normal (see Figure~\ref{fig:kdensity_panels}) so that inferences based on first-order asymptotic results are unreliable. Because trees are sensitive to hyperparameter choice (tree depth, etc.), we experimented with an alternative strategy for hyperparameter tuning (described in Online Supplement~\ref{sec:simresults}) that led to normal sampling distributions but larger biases. An extensive discussion on hyperparameter tuning can be found in Section~\ref{sec:tune_mc} of the Online Supplement.

The superior performance of LASSO is indicated by the Root Mean Square Errors (RMSEs) of the estimator, model, and nuisance parameters. These results support the use of an \emph{ensemble} learning strategy, in which the best-performing learner is chosen because, for our example, this would have selected LASSO and ensured reliable inference.  

\input{tabs/06_mc_simul}

There is little to choose between the CRE, FD (Exact) and WG (Approximate) procedures.  This is not unexpected because the DGP in the simulation study satisfies the additional assumptions for modelling the unobserved heterogeneity with Mundlak-like CRE, but the results are slightly different because these procedures involve learning different nuisance functions. More generally, the FD (Exact) estimator is the most robust procedure because it does not rely on any additional assumptions.  In comparison, the WG (Approximation) method performed poorly no matter which learner was chosen because the approximation error induced by using the linearly transformed data (the time-demanded transformation) is large for the non-linear DGP.  

%% file: tabs/06_mc_simul.tex
\begin{table}[t!]
\centering
 \caption{Average MC simulation results, nonlinear and discontinuous DGP}\label{tab:mx_collapse}
\scalebox{.9}{
\begin{threeparttable}	
\begin{tabular}{lcccccc}  
\vspace{-3mm}\\   
\hline\hline
\vspace{-3mm}\\
&	Bias$(\thetano$)&	RMSE$(\thetano)$&	SE$(\thetano)$/SD$(\thetano)$& Model RMSE& 	RMSE$_l$& RMSE$_m$\\
\hline

\vspace{-1mm}\\
\multicolumn{7}{l}{\emph{Panel A: CRE approach}}\\
\vspace{-3mm}\\
OLS&0.993	&	0.993	&	0.999&&&\\
DML-Lasso	&	0.009	&	0.014	&	1.235	&	5.818	&	1.981	&	1.432	\\
DML-CART	&	-0.087	&	0.199	&	0.084	&	21.316	&	7.169	&	5.910	\\
DML-RF	&	0.149	&	0.151	&	1.320	&	6.773	&	2.427	&	1.779	\\
DML-Boosting	&	-0.007	&	0.033	&	0.871	&	7.432	&	2.523	&	1.860	\\

\vspace{-1mm}\\
\multicolumn{7}{l}{\emph{Panel B: Exact approach with FD transformation}}\\
\vspace{-3mm}\\
OLS&0.993	&	0.993	&	0.951&&&\\
DML-Lasso	&	0.005	&	0.008	&	1.050	&	4.302	&	1.605	&	1.432	\\
DML-CART	&	0.291	&	0.385	&	0.049	&	46.564	&	18.889	&	12.596	\\
DML-RF	&	0.752	&	0.765	&	0.071	&	14.180	&	8.824	&	5.840	\\
DML-Boosting	&	-0.014	&	0.059	&	0.461	&	9.815	&	3.514	&	2.612	\\

\vspace{-1mm}\\
\multicolumn{7}{l}{\emph{Panel C: Approximation approach for WG transformation}}\\
\vspace{-3mm}\\
OLS&0.993	&	0.993	&	0.999&&&\\
DML-Lasso	&	0.977	&	0.977	&	1.120	&	4.347	&	9.625	&	6.450	\\
DML-CART	&	0.754	&	0.764	&	0.050	&	20.049	&	11.363	&	7.552	\\
DML-RF	&	0.972	&	0.972	&	0.739	&	5.070	&	9.817	&	6.578	\\
DML-Boosting	&	0.918	&	0.918	&	0.426	&	9.837	&	10.017	&	6.711	\\
\hline
\end{tabular} 	
\footnotesize
\renewcommand{\baselineskip}{11pt}
\textbf{Note:} The figures in the table are the average values over the total number of Monte Carlo replications ($R=100$). The true target parameter is $\theta = 0.50$; $N=4000$ and $T=10$. %
The quantities displayed correspond to: $Bias(\thetano, \theta) = R\inv\sum_{r=1}^R\big(\E(\thetano_r) - \theta\big)$; $RMSE(\thetano) = \sqrt{R\inv\sum_{r=1}^R\big(Var(\thetano_r)  + Bias(\thetano_r, \theta)^2\big)}$, where $Var(\thetano) = \E\big[\big(\thetano - \E(\thetano)\big)^2\big]$; Model~RMSE$=(RK)\inv\sum_{r=1}^R\sum_{k=1}^K\sqrt{(|W|)\inv \sum_{i\in W} \big(\widehat{\uit} - \widehat{\vit} \thetano_k)^2 }$, where $\widehat{\uit}$ and $\widehat{\vit}$ are the residuals of the PO structural equations; the RMSE of the nuisance parameters are $RMSE_l = (RK)\inv\sum_{r=1}^R\sum_{k=1}^K\sqrt{(|W^c|)\inv  \sum_{i\in W^c} \big(\yit - \widehat{l}_k(.)\big)^2 }$ and $RMSE_m =  (RK)\inv\sum_{r=1}^R\sum_{k=1}^K\sqrt{(|W^c|)\inv \sum_{i\in W^c} \big(\dit - \widehat{m}_k(.)\big)^2 }$.
DML-Lasso uses 525 raw variables; the rest of the learners 30 raw variables.  We use the Neyman-orthogonal PO score and five-fold cross-fitting.
\end{threeparttable}
}
\end{table}

%% file: 05_application.tex
\section{Empirical Application}\label{sec:application}
\setcounter{equation}{0}

We reanalyse \cite{fazio2023}'s study of the impact of the National Minimum Wage (NMW) on voting behaviour in the UK.  The data used in the original investigation come from the British Household Panel Survey (BHPS) comprising $4,927$ working individuals (aged 18-64) between waves 1 and 16 (from 1991 until 2007).\footnote{BHPS is a longitudinal survey study for British households that run from 1991 until 2009. The online replication package provides the instructions for data access and the codes to run the analysis in this section.  DML estimation is conducted in \textsc{R} using the \texttt{XTDML} package in the replication package (or accessible in its latest version at \url{https://github.com/POLSEAN/XTDML} at the time of writing)  which is built on \texttt{DoubleML} by \citet{bach2024}.   \nocite{bhps}} 
The treatment is measured by the question \emph{`Were you paid the minimum wage'} asked to those responding at Wave 9 in 1999. The authors use least squares to estimate the average treatment effect on the treated (ATT) between the NMW policy on various outcomes, including voting for conservative political parties which is our outcome of interest. 
\cite{fazio2023} compared the results of four regression specifications with different sets of covariates and fixed effects to capture all potential confounders, showing that the estimates did not vary across these, hence, concluding that workers who were paid the NMW rate were more likely to support conservative parties. 

We revisit Specification~(2) of Table~5 from the original paper that includes all control variables, regions and wave fixed effects, by using DML procedure to fit the following PO-PLPR model
\begin{align}
     Vote_{it} & = \vit\theta_0 + l_1(\vector  X_{it})+\alpha_i+\uit\\
     \vit & = NMW_{it} - m_1(\vector  X_{it})-\gamma_i
\end{align}
\noindent where $Vote_{it}$ is an indicator corresponding to one if respondent $i$ voted for conservative parties in wave $t$, and zero otherwise; $NMW_{it}$ is the treatment variable which is equal to one if the respondent's hourly pay increased due to the introduction of the NMW in 1999 and the respondent is observed in wave 9 onwards; $\vector  X_{it}$ are the confounding variables whose functional form is \emph{ex ante} unknown; and $\theta$ is the target parameter.  The original base control variables included information about the age, education, marital status, household size, and income of other household members for respondent $i$ in wave~$t$. 

We add thirty additional variables to the original specification, providing information on demographic characteristics, socio-economic status, employment and work-related variables, and ideology of the respondents. The complete list of confounders is reported in Table~\ref{tab:appx_descriptive} of the Online Supplement~\ref{sec:appx_var_des}. We also increase the sample size by including two additional waves for years 2008 and 2009 (waves 17 and 18, respectively). Our final sample includes $9,922$ working individuals.
The functional form of the confounding variables is learnt using four different base learners:\ LASSO with a dictionary of non-linear terms (i.e., polynomials of order three and interaction terms of each raw variable), CART, RF, and boosted trees. The hyperparameter tuning of the base learners used in the implementation is discussed in the Online Supplement~\ref{sec:tune_appl}.

\input{tabs/07_tab_empirical_application}
The point estimates of NMW effect on voting conservative parties are reported in Table~\ref{tab:fazio5}. Column~(1) shows OLS estimates (i.e.\ using standard panel estimation based on linear models) while the remaining columns contain the results using DML with different learners. Panel A displays the CRE estimates, Panel B the FD (Exact) estimates, and Panel C the WG (Approximation) estimates.
The regression equations include the individual means of all included control variables when the CRE is used, and one-period lagged variables of the controls when FD exact is used. Cluster-robust standard errors are reported in parenthesis with clustering at the respondent level.

The results exhibit considerable differences in the estimated effects between learners and estimators but, we argue, are consistent with those from a DGP which is either close to linear or smoothly non-linear as in our simulation study (see Online Supplement~\ref{sec:appx_mc_simul}). 
The first difference we consider is between the FD (Exact) and CRE estimators. The former should be the most robust because it does not rely on the Mundlak-type models for fixed effects and also allows for different sets of omitted variables fixed only between distinct wave pairs $t-1$ and $t$ (which could be represented by adding parameters like $\alpha_{i(t-1,t)}$ to the models). The magnitudes of the FD (Exact) point estimates are (a) smaller for OLS and LASSO and (b) more stable across all learners than those based on CRE. The RMSEs for the outcome ($l$) and treatment ($m$) models for FD (Exact) are smaller than those for CRE, which indicates the learners are more effective at learning nuisance functions of $\vector  X_{it}$ and $\x_{it-1}$ than of $\vector  X_{it}$ and $\overline{\x}_i$ and so more likely to have errors bounded by $N^{1/4}$. Moreover, the consistency across OLS and the learners would suggest a DGP that is linear or smoothly non-linear. 
Second, the WG (Approximate) estimates are also consistent across learners and larger than those obtained using FD (Exact). The differences between FD (Exact) and WG (Approximate) are also commensurate with the additional robustness of the former. 
The CRE estimates obtained using tree-based learners, compared with all other estimates, are both larger (in absolute value) and less precise, with the DML-Boosting estimate particularly imprecise. However, the RMSEs indicate the tree-based algorithms do a better job learning the nuisance functions in $\vector  X_{it}$ and ${\bar {\bf x}}_i$ than LASSO, and an ensemble strategy would have selected the estimate based on the random forest (DML-RF) for comparison with OLS and so pointed to a slightly larger but less significant estimate than the best estimate obtained using WG (Approximate). 
We finally reiterate the preference of FD (Exact) in this application, and the importance of comparing different procedures and learners in reaching this conclusion.

%% file: tabs/07_tab_empirical_application.tex
\begin{table}[t!]
\centering
 \caption{The Effect of National Minimum Wage on Voting Behaviour in the UK}\label{tab:fazio5}
\scalebox{.9}{
\begin{threeparttable}	
  \begin{tabular}{lccccc}   
\vspace{-3mm}\\   
\hline\hline
\vspace{-3mm}\\	
& OLS&  DML-Lasso& DML-CART& DML-RF& DML-Boosting\\
&(1)&(2)&(3)&(4)&(5)\\
\vspace{-3mm}\\
\hline

\multicolumn{6}{c}{\emph{Dependent variable: Voting for conservative parties}}\\

\vspace{-2mm}\\
\multicolumn{6}{l}{\emph{Panel A: CRE approach}}\\
\vspace{-3mm}\\
NMW &0.051***&0.048**&0.069*&0.180&-0.319\\
&(0.019)&(0.019)&(0.036)&(0.151)&(0.278)\\
\vspace{-3mm}\\
RMSE of learner $l$&& 0.4574&0.4195&0.4130&0.4350\\
RMSE of learner $m$&&0.0645&0.0512&0.0155&0.0107\\
Model RMSE&&1.1143& 1.030&1.0085&1.0478\\
No. control variables&72&1,476&72&72&72\\
\hline

\vspace{-2mm}\\
\multicolumn{6}{l}{\emph{Panel B: Exact approach with FD}}\\
\vspace{-3mm}\\
$\Delta$NMW&0.022*&0.021&0.026&0.018&0.024\\
&(0.013)&(0.026)&(0.026)&(0.026)&(0.026)\\
\vspace{-3mm}\\
RMSE of learner $l$&&0.2821&0.2823&0.2815&0.2859\\
RMSE of learner $m$&&0.0504&0.0499&0.0492&0.0533\\
Model RMSE&&0.6419&0.6427&0.6404&0.6261\\
No. control variables&72&1,476&72&72&72\\

\hline
\vspace{-2mm}\\
\multicolumn{6}{l}{\emph{Panel C: Approximation approach with WG}}\\
\vspace{-3mm}\\
$\widetilde{\rm{NMW}}$&0.051***&0.046**&0.048**&0.040**&0.048***\\
&(0.019)&(0.019)&(0.019)&(0.018)&(0.017)\\
\vspace{-3mm}\\
RMSE of learner $l$&&0.2121&0.2130&0.2124&0.2155\\
RMSE of learner $m$&&0.0642&0.0645&0.0651&0.0711\\
Model RMSE&&0.5276&0.5299&0.5278&0.5239\\
No. control variables&36&738&36&36&36\\



\hline
\end{tabular} 	
\footnotesize
\renewcommand{\baselineskip}{11pt}
\textbf{Note:} The table displays our estimates based on Specification~(2) of Table~5 in \cite{fazio2023} using a different sample and confounders. Figures in Column~(1) are estimated through OLS while the remaining columns using DML with different learners. 
The number of observations ($NT$) is $59,745$ in Panels A, C and D while $49,823$  in Panel B; the number of cross-sectional units is $9,922$ in  all panels. 
The DML estimation uses 5-fold block cross-fitting and partialling-out score. The hyperparameters of the base learners are tuned with grid search.
Cluster-robust standard errors at the respondent level in parenthesis. Significance levels: * p $<$ 0.10, ** p $<$ 0.05, *** p $<$ 0.01.
\end{threeparttable}}
\end{table}


%% file: 06_conclusion.tex
\section{Conclusion}\label{sec:conclusion}
DML is a powerful tool for leveraging the power of ML for robust estimation of treatment effects, or policy-intervention. 
We provide estimation tools for applying DML when panel data are available which practitioners can use in place of existing ones or in a complementary way to test the robustness of their results to non-linearity. Moreover, the nature of static panels means that our procedures extend naturally to unbalanced panels provided that the non-response at each wave is conditionally independent of $\Yit$ given $\Xit$ and $\xi_i$. Further work on relaxing this assumption for CRE could build on \mbox{\cite{wooldridge2019}.}

In general, although the three approaches we presented can be considered complementary to each other, the suitability of each will depend on the specific empirical example and the assumptions that the analyst is prepared to make. However, following our empirical studies, our recommendation for practice is to employ the FD (exact) approach because the ML algorithms we consider proved effective at estimating the nuisance functions involving predictors from waves $t$ and $t-1$, the random coefficient treatment effect heterogeneity of \cite{wooldridge2019} is automatically accounted for, and it places fewer assumptions on the the distribution of the fixed effects. 



From our simulation study, we also recommend using multiple (base) learners for DML stage one in the context of \emph{ensemble} learning. In our simulations, the LASSO with an extended dictionary of non-linear terms performed best across all scenarios. However, despite Lipschitz continuity and weak sparsity conditions holding, we found tree-based learners to perform poorly in terms of bias (of the point and interval estimates) and in terms of standard deviation and normality, even after making extensive efforts to implement adaptive hyperparameter tuning.  We do not claim this to be a general result (there are many successful applications of tree-based learners in the literature), but we strongly recommend that analysts use an \emph{ensemble} strategy involving multiple learners (the tree-based learners could perform better than the others in some scenarios) as is standard practice in other disciplines where ML and other AI-based methods are widely used. 

Finally, our method is limited when there is treatment effect heterogeneity but the analyst still wishes to target the ATE rather than the CATE.  Further work on extending these procedures, based on adapting the {\em interactive model} of \citet[Sec.\ 5]{chernozhukov2018} to panel data, would therefore be of great value for practice.

%% file: 888_proof_prop1.tex
\section{Derivations of Results}
\setcounter{equation}{0}

\medskip
\subsection{Derivation of learning procedure for CRE in Section~5.1.1}\label{sec:proof_lemma_cre}
If $\bvi\approx 0$ and $D_{i1},\dots,D_{iT}|\Xii,\overline{\vector X}_i\distas{}{} N(\bm{\mu}_d, \bm{\Sigma}_d)$,  where
\begin{equation}
\bm{\mu}_d = 
\begin{pmatrix}
 \tm_1(\vector X_{i1},\overline{\vector X}_i) \\
 \hdots \\
  \tm_1(\vector X_{iT},\overline{\vector X}_i) \\
\end{pmatrix}
 \hspace{0.2cm} {\rm and} \hspace{0.2cm} 
\bm{\Sigma}_d = %
\begin{pmatrix}
    \sigma_v^2+\sigma_c^2& \dots& \sigma_c^2\\
    \vdots& \ddots & \vdots\\
    \sigma_c^2& \dots& \sigma_v^2+\sigma_c^2
\end{pmatrix}
\end{equation}
\noindent then elementary calculations for multivariate normal distributions can be used to derive the joint distribution $D_{i1},\dots,D_{iT}|\Xii,\overline{\vector X}_i$ from which it can be shown that
\begin{equation}
  E[\Dit|\Xii,\overline{\vector X}_i,\bdi]=\widetilde{m}_1(\vector X_{it},\overline{\vector X}_i)+\bdi-\bmi_1(\Xii,\overline{\vector X}_i)\equiv m_1^*(\vector X_{it},\overline{\vector X}_i,\bdi),
\end{equation}
\noindent where $\bmi_1(\Xii,\overline{\vector X}_i)=T\inv\sum_{t=1}^T\tmi_1(\vector X_{it},\overline{\vector X}_i)$ 
constrains \mbox{$T\inv \sum_{t=1}^T m^*_1(\vector X_{it},\overline{\vector X}_i,\bdi) = \bdi$.} This constraint can be treated as implicit to be picked up by the learner from  $\{\Dit,\vector X_{it},\overline{\vector X}_i:t=1,\dots,T\}_{i=1}^N$ (supervised by $\Dit$).

The general approach in non-normal cases is simply to construct $m_1^*(\Xii,\overline{\vector X}_i,\bdi)$ by supervised learning of $\tmi_1(\vector X_{it},\overline{\vector X}_i)$ from $\{\Dit,\Xii,\overline{\vector X}_i:t=1,\dots,T\}_{i=1}^N$ and $\bmi_1(\Xii,\overline{\vector X}_i)$ calculated directly from these estimates combine to give $\tmi_1(\vector X_{it},\overline{\vector X}_i)+\big\{\bdi-\bmi_1(\Xii,\overline{\vector X}_i)\big\}=\tmi(\vector X_{it},\overline{\vector X}_i)+c_i$ as required.

\medskip
\subsection{Justification of Approximation Approach in Section~5.1.2.a} \label{sec:proof_lemma_approx}

A first-order Taylor series expansion of $l_1$ around some fixed value $\x$ gives $l_1(\vector X_{it})=(\vector X_{it}-\x)\dot{l}_1(\x)+\mathcal{O}(\|\vector X_{it}-\x\|^2)$, where $\|.\|$ is the $L_1$-norm and column-vector $\dot{l}_1(\x)$ is the partial derivative of $l_1$ with respect to $\vector X_{it}$ evaluated at $\x$, so that 
$Q(l_1(\vector X_{it}))=Q(\vector X_{it})\dot{l}_0(\x)+\mathcal{O}(\|\mathbf{b}_x\|^2)$, where $\mathbf{b}_x={\rm sup}_t\|\vector X_{it}-\x\|$. Hence, there is some $Q(\vector X_{it})\dot{l}_{0}(\overline{\x})+\mathcal{O}\big(\|\mathbf{b}\|^2\big)\approx Q(\vector X_{it})\dot{l}_{0}(\overline{\x})$ minimizing some loss function, where $\mathbf{b}=\inf_{\x}\mathbf{b}_{\x}$ and $\overline{\x}=\arginf_{\x}\mathbf{b}_{\x}$ are respectively the smallest bound over the bounded support of all possible $\x$-centered confounders, and $\overline{\x}$ is any value obtaining this bound. If $l_1$ is continuous and differentiable over the support (e.g.\ it is Lipschitz continuous), the mean-value inequality gives that $\overline{\x}$ exists, but whether or not it is a good approximation remains to be determined empirically.

\medskip
\subsection{Derivation of Exact Approach in Section~5.1.2.b}\label{sec:proof_lemma_exact}

The key step is showing that $E[\Yit-Y_{it-1} \mid \vector X_{it-1},\vector X_{it},\xi_i] =E[\Yit | \vector X_{it}, \xi_i]-E[Y_{it-1}\mid \vector X_{it-1},\xi_i]$.  This follows under Assumptions~\ref{item:asm_feedback}-\ref{item:asm_nolags} where the DGP factorizes as
\begin{equation*}
{\cal P}(\xi_i)
\prod_{t=1}^T {\cal P}\big(\vector X_{it} | L_t(\vector X_i),\xi_i\big){\cal P}(D_{it},y_{it}|\vector X_{it},\xi_i) = p\{L_T(\vector X_i)\}p\{\xi_i | L_T(\vector X_i\}\} \prod_{t=1}^T{\cal P}(\Dit,\Yit | {\bf x}_{it},\xi_i),
\end{equation*}
where we distinguish between components of the DGP ${\cal P}(.)$ and general densities $p(.)$. As such,
\begin{equation}
E[\Yit\mid\vector X_{it-1},\vector X_{it},\xi_i] =E_{\cal P}[E_{\cal P}[\Yit\mid\Dit,\vector X_{it},\xi_i]\mid \vector X_{it},\xi_i]=l_1(\vector X_{it})+\alpha_i,
\end{equation}
and without loss of generality, for treatments with discrete support,
\begin{align*}
&E[Y_{it-1} |\vector X_{it-1},\vector X_{it},\xi_i] \notag \\  
&= \int_{Y_{it-1}}Y_{it-1} \sum_{D_{it-1},\Dit} {\cal P}(D_{it-1}\mid \vector X_{it-1},\xi_i){\cal P}(Y_{it-1} \mid D_{it-1},\vector X_{it-1},\xi_i){\cal P}(\Dit|\vector X_{it},\xi_i) d{F(Y_{it-1})} \\
&=E[Y_{it-1} \mid \vector X_{it-1},\xi_i]=l_1(\vector X_{it-1})+\alpha_i,
\end{align*}
where $F$ is the distribution function for $Y_{it-1}$. Hence, 
\begin{align}
E[\Yit-Y_{it-1} \mid \vector X_{it-1},\vector X_{it}] &= E[E[\Yit-Y_{it-1} \mid \vector X_{it-1},\vector X_{it},\xi_i]\mid\vector X_{it},\vector X_{it-1}] \\
& =E[l_1(\vector X_{it})+\alpha_i-l_1(\vector X_{it-1})-\alpha_i\mid\vector X_{it},\vector X_{it-1}] \\ & =l_1(\vector X_{it})-l_1(\vector X_{it-1}) \notag
\end{align}
as required.  The same approach can be used to show that $E[\Dit - D_{it-1}\mid\vector X_{it}, \vector X_{it-1},\xi_i]=m_1(\vector X_{it})-m_1(\vector X_{it-1})$.

Under Assumptions~\ref{item:asm_feedback}-\ref{item:asm_additive}, it follows that $E[\Yit|\vector X_{it},\alpha_i]=E[\Yit\mid\vector X_{it},\overline{\vector X}_i,a_i]$ and $E[\Dit|\vector X_{it},\gamma_i]=E[\Dit|\vector X_{it}, \overline{\vector X}_i,c_i]$.  Then, from the assumptions required to derive the CRE model in Section~\ref{sec:three_approaches}, $E[\Yit|\vector X_{it},\overline{\vector X}_i,a_i]=l_1(\vector X_{it})+\omega_\alpha(\overline{\vector X}_i)+a_i$ and $E[\Dit|\vector X_{it},\overline{\vector X}_i,c_i]=m_1(\vector X_{it})+\omega_\gamma(\overline{\vector X}_i)+c_i$.  Further, by iterated expectations with respect to the data generating process $p(\alpha_i|X_i)=p(\alpha_i|\overline{\vector X}_i)$, this implies that $E[\Yit|\vector X_{it},\overline{\vector X}_i]=l_1(\vector X_{it})+\omega_\alpha(\overline{\vector X}_i)+E[a_i|\overline{\vector X}_i]\equiv\tl_1(\vector X_{it},\overline{\vector X}_i)$ and $E[\Dit|\vector X_{it},\overline{\vector X}_i]=m_1(\vector X_{it})+\omega_\gamma(\overline{\vector X}_i)+E[c_i|\overline{\vector X}_i]\equiv\tm_1 (\vector X_{it},\overline{\vector X}_i)$ so that $Q\big(\tl_1(\vector X_{it},\overline{\vector X}_i)\big)=Q\big(l_1(\vector X_{it})\big)$ and $Q\big(\tm_1 (\vector X_{it},\overline{\vector X}_i)\big)=Q\big(m_1(\vector X_{it})\big)$. 

%% file: 888_score.tex
\section{Derivation of Neyman Orthogonal Score Function}\label{sec:proof:score}
\setcounter{equation}{0}

Generally, suppose that nuisance parameter $\bmeta$ is a finite-vector of square-integrable and Lipschitz-continuous functionals where true value $\bmeta_0\in {\cal T}$ a convex subset of a normed vector space.  
The Neyman-orthogonal score $\psi^\perp(W_{it};\btheta,\bmeta)$ satisfies (a) moment condition $E[\psi^\perp(W_{it};\btheta_0,\bmeta_0)]=0$ for unique $\btheta_0$, and (b) $\partial_{\eta} E\big[\psi^\perp(W_{it};\btheta_0,\bmeta_0)\big][\bmeta-\bmeta_0]=0$, where expectations are with respect to the data generating process ${\cal P}$ for random vector $W_{it}$.

The Neyman-orthogonal score presented in Section~\ref{sec:no_score} adapts the development outlined for the cross-sectional PLR model by \citet[Section\ 2.2.4]{chernozhukov2018} for the PO-PLRP model \eqref{eqn:plr_y}-\eqref{eqn:plr_v}. 
To derive it, we begin by considering a more general version of PLRP (\ref{eqn:plriv}) where Assumption~\ref{item:asm_effect_fe} is relaxed to allow for heterogeneity along the lines set out in Section \ref{sec:hetero} when $E[\alpha^*_i\mid \Dit,\vector{X}_{it}]=0$, that is, $\alpha^*_i$ is a random effect rather than a fixed effect:
\begin{equation*}
    \Yit=\Dit {\bf S}_{it}^\prime{\pmb\theta}_0+g_1(\vector X_{it})+r_{it},
\end{equation*}
\noindent where $r_{it}=\alpha^*_i+\uit$ and ${\bf S}_{it}=S(\Xit,t)$ is the user-specified heterogeneity model that can include $t$ to allow parameters to vary by wave.\footnote{The linear part could be extended without loss of generality to include functions of $t$ and $\vector X_{it}$ alone (i.e.\ no interactions with $\Dit$) were the analyst to decide to decompose $g_1$ into non-parametric and linear-parametric parts with additively separable effects.} 


This model can be written vector-wise as
\begin{equation}
    {\bf r}_i=\vector Y_i-\mathbf{f}_{\theta_0}(\vector D_i,X_i)-\gi_1(X_i),
\end{equation}
\noindent where ${\bf r}_i=(r_{i1},\ldots,r_{iT})'$, $\vector Y_i=(Y_{i1},\ldots,Y_{iT})'$,
\begin{equation}
   \mathbf{f}_{\theta_0}'(\vector D_i,X_i)=\big(D_{i1}{\bf S}_{i1}^\prime\pmb{\theta}_0,\dots, D_{iT}{\bf S}_{iT}^\prime\pmb{\theta}_0\big)={\rm diag}({\bf D}_i)S_i^\prime\btheta_0,
\end{equation}
$\gi_1(X_i)=\big(g_1(\vector X_{i1}),\dots, g_1(\vector X_{iT})\big)'$ and $X_i=\{\vector X_{it} : t=1,\dots,T\}$ a set of time-varying predictors; $S_i=({\bf S}_{i1},\ldots,{\bf S}_{iT})$ and ${\rm diag}({\bf D}_i)$ is a $T\times T$ diagonal matrix with diagonal elements ${\bf D}_i$.

By construction, the conditional moment restriction $E[\uit|\Dit,\vector X_{it},\alpha_i^*]=0$ holds, but further assumptions are generally needed to identify $\theta_0\equiv\pmb{\theta}_0$.  
From the random effects assumption, it follows that
\begin{equation}
\label{eqn:eqn0}
    E[{\bf r}_i\mid\vector D_i,X_i]=\zero.
\end{equation}
and from the remarks following \citet[Definition 2.6]{chernozhukov2018}, Neyman orthogonality is demonstrated by showing the partial derivative of the score function, treating each functional component of $\bmeta$ as a scalar parameter, with respect to $\bmeta$ is zero at $(\theta_0,\bmeta_0)$. 

Converting the notation used in \citet[Section\ 2.2.4]{chernozhukov2018} to that used in this paper, $W\equiv\{\vector Y_i,\vector D_i\}\cup X_i$, $R\equiv\{\vector D_i\}\cup X_i$ and $Z=X_i$, with $h(Z)\equiv\gi_1(X_i)$ and $m(W;\theta,h(Z))\equiv{\bf r}_i$.  Taking $\partial_{\theta}\equiv\partial/\partial\btheta=\nabla_{\btheta}$ to be the partial derivative, the first component of their Lemma~2.6 (p.~C20) is
\begin{equation*}
    A(R)\equiv -\partial_{\theta'}E\big[m\{W;\theta,h_0(Z)\}\mid R\big]\mid_{\theta=\theta_0}=-\partial_{\btheta} \mathbf{f}_{\theta_0}'=-S_i{\rm diag}({\bf D}_i),
\end{equation*}
\noindent which equals $-\vector D_i$ if $f_{\theta_0}(\vector X_{it},t)=\theta_0$ under Assumption~\ref{item:asm_effect_fe}.

The next component is 
\begin{equation*}
 \Gamma(R)\equiv-\partial_{\nu'}\E\big\{m(W;\theta_0,\nu)|R\big\}|_{\nu=h_0(Z)}=-I_T,   
\end{equation*}
\noindent that is, the $T\times T$ identity matrix; this is based on $h(Z)= (g_{11},\ldots,g_{1T})'$, that is, treating the nuisance functions at each wave as distinct, but the final result is the same if we constrain $h(Z)\equiv g_1$ to give $\Gamma(R) = {\bf 1}_T$ a column vector of ones.  

Lastly,
\begin{equation*}
  \Omega(R)\equiv E\big[m\{W;\theta_0,h_0(Z)\}m'\{W;\theta_0,h_0(Z)\}\mid R\big]=E[{\bf r}_i{\bf r}_i'|\vector D_i,X_i]=\Sigma_0(\vector D_i,X_i),  
\end{equation*}
\noindent that is, the $T\times T$ within-individual auto-covariance matrix; and
\begin{align*}
G(Z) & \equiv E[A'(R)\Omega\inv(R)\Gamma(R)\mid Z]\big\{ \Gamma'(R)\Omega\inv (R)\Gamma(R)|Z\big\}\inv  \\
& = %
E[\partial_{\btheta} \mathbf{f}_{\theta_0}'\Sigma_0\inv(\vector D_i,X_i) I_T |X_i] E[\I_T \Sigma_0\inv(\vector D_i,X_i) \I_T|X_i]\inv \\
& =E[\partial_{\btheta} \mathbf{f}_{\theta_0}' \Sigma_0\inv(\vector D_i,X_i)| X_i] E[\Sigma_0\inv(\vector D_i,X_i)|X_i]\inv.
\end{align*}

\noindent Then from \citet[equation (2.23)]{chernozhukov2018},
\begin{align*}
\mu(R) %
    & \equiv A'(R)\Omega\inv(R) - G(Z)\Gamma'(R)\Omega\inv(R) \\
    & =\partial_{\btheta} \mathbf{f}_{\theta_0}' \Sigma_0\inv(\vector D_i,X_i)-E[\partial_{\btheta} \mathbf{f}_{\theta_0}' \Sigma_0\inv\mid X_i]E[\Sigma_0\inv\mid X_i\big]\inv \Sigma_0\inv\\
    &=\big[\partial_{\btheta} \mathbf{f}_{\theta_0}'-E[\partial_{\btheta} \mathbf{f}_{\theta_0}'\Sigma_0\inv\mid X_i]E[\Sigma_0\inv|X_i]\inv \big]\Sigma_0\inv,
\end{align*}
where $\Sigma_0=\Sigma_0(\vector D_i,X_i)$. 

Hence, the Neyman orthogonal score $\psi^\perp(W;\theta,h(Z))\equiv\mu(R)m\{W;\theta,h(Z)\}$ is
\begin{equation}\label{eqn:score1}
\psi^\perp(W;\theta_0,h_0(Z))=\Big\{\partial_{\btheta} \mathbf{f}_{\theta_0}'-E[\partial_{\btheta} \mathbf{f}_{\theta_0}' \Sigma_0\inv\mid X_i]E[\Sigma_0\inv|X_i]\inv\Big\}\Sigma_0\inv{\bf r}_i.    
\end{equation}
Finally, by exploiting that ${\bf r}_i\ {\perp \!\!\! \perp}\ \vector D_i \mid X_i$, score~\eqref{eqn:score1} simplifies as
\begin{equation}\label{eqn:score2}
 \psi^\perp(W;\theta_0,h_0(Z))=\big\{\partial_{\btheta} \mathbf{f}_{\theta_0}'-E[\partial_{\btheta} \mathbf{f}_{\theta_0}'\mid X_i]\big\} \Sigma_0\inv( X_i){\bf r}_i
\end{equation}
because $\Sigma_0(\vector D_i,X_i)=\Sigma_0(X_i)$; and, under Assumption~\ref{item:asm_effect_fe}, $f_{\theta_0}(\vector X_{it},t)=\theta_0$ so that
\begin{equation*}
    \psi^\perp(W;\theta_0,h_0(Z))=\{\vector D_i-{\bf m}_1(X_i)\} \Sigma_0\inv(X_i) {\bf r}_i,
\end{equation*}
\noindent where ${\bf m}_1(X_i)=(m_1({\vector X}_{i1}),\ldots,m_1(\vector X_{iT}))^\prime$.

We now show that an equivalent result is obtained for PO-PLRP model~\eqref{eqn:plr_y}-\eqref{eqn:plr_v} under random effects assumption $E[\alpha_i\mid {\bf D}_i,X_i]=0$.  The model is
\[
{\bf r}_i=\vector Y_i-\mathbf{f}_{\theta_0}(\vector V_i,X_i)-{\bf l}_1(X_i),
\]
\noindent where ${\bf V}_i={\bf D}_i-{\bf m}_1(X_i)$, 
${\bf l}_1(X_i)=(l_1(\vector X_{i1}),\dots, l_1(\vector X_{iT}))^\prime$, $r_{it}=\alpha_i+\uit$ and the heterogeneity function
\[
\mathbf{f}_{\theta_0}(\vector V_i,X_i)={\rm diag}({\bf V}_i)S_i^\prime\pmb{\theta}_0.
\]

Here, $h(Z)\equiv ({\bf l}_1,{\bf m}_1)$, $A(R)\equiv -S_i{\rm diag}({\bf V}_i)$, 
$\Omega(R)=\Omega(Z)$ and $\Gamma(R)=\Gamma(Z)\equiv\big(-I_T, {\rm diag}({\bf f}_{\theta_0})\big)$, where ${\rm diag}({\bf f}_{\theta_0})$ is a $T\times T$ matrix with diagonal elements given by ${\bf f}_{\theta_0}$ with all off-diagonal elements equal to zero, under which
\[
G(Z)=E\big[A\prime(R)|Z\big]\Omega(Z)\inv\Gamma(Z)\big\{\Gamma'(Z)\Omega\inv(Z)\Gamma(Z)\big\}\inv={\bf 0}
\]
\noindent and $\mu(R)=A\prime(R)\Omega\inv(Z)$.

These results apply to the fixed-effects PO-PLRP procedures from Section~\ref{sec:three_approaches} as follows. For the CRE procedure, $W\equiv\{\vector Y_i,\vector D_i, \overline{\vector X}_i\}\cup X_i$, $R\equiv\{\vector D_i,\overline{\vector X}_i\}\cup X_i$ and $Z\equiv\{\overline{\vector X}_i\}\cup X_i$, with $h(Z)\equiv ({\widetilde{\bf l}}_1,{\widetilde{\bf m}}_1)$ and $m(W;\theta_0,h_0(Z))\equiv{\bf r}_i$ where $R_{it}=a_i+\uit$. Note that from Section~\ref{sec:learning_cre} the required form of nuisance function is shown to be ${\widetilde{\bf m}}_1=\{{\widetilde m}_1(\vector X_{it},\vector \bxi)+c_i\}_{t=1}^T$.

For the transformation approaches, $W\equiv\{Q(\vector Y_i),\vector D_i\}\cup X_i$, $R\equiv\{\vector D_i\}\cup X_i$ and $Z\equiv X_i$, with $m(W;\theta_0,h_0(Z))\equiv{\bf r}_i$ where $R_{it}=Q(\uit)$ and $Q(\vector Y_i)=\{ Q(\Yit)\}_{t=1}^T$. The treatment function is $Q(\vit {\bf S}^\prime_{it})\pmb{\theta}_0$ (simplifying as 
$Q(\vit)\theta_0$ under Assumption~\ref{item:asm_effect_fe}) and $R_{it}=Q(\uit)$ does not depend on $\alpha_i$. The nuisance functions $h(Z)$ appear linearly on the right-hand side of $r_{it}$ as $-Q(l_1(\vector X_{it}))$ and $Q(m_1(\vector X_{it}){\bf S}_{it}^\prime)\pmb{\theta}_0$ (or $Q(m_1(\vector X_{it}))\theta_0$ under homogeneity) and so the form of the estimating equation is the same with the orthogonalized regressor equal to $Q(\vit S_{it})$ (or $Q(\vit)$ for homogeneous treatments).

The full set of conditions required for our semi-parametric model to be a Neyman orthogonal score are set out in Lemma 2.6 of \citet{chernozhukov2018}:\ their condition (a) holds under our model assumptions; condition (b) reduces to $||E[\partial_{\theta}f_{\theta}\mid X_i]||^4,||\partial_{\theta}f_{\theta}||^2$ and $||\Sigma(X_i)||^{-2}$ all having finite expectations under the data generating process; and condition (c) is that for any $h\in{\cal T}$ such that $h\neq h_0$ ($h_0$ is the true nuisance function) there exists finite $C_h$ such that $\Pr(|| E[\psi^\perp | {\bf D}_i,X_i] || < C_h) = 1$.


%% file: SI_mc_simulations.tex
\section{Monte Carlo Simulations}\label{sec:appx_mc_simul}
\subsection{Monte Carlo Simulation Design}\label{sec:mc_design}
We generate data for the Monte Carlo simulations from the data generating process (DGP) below
\begin{align}
   &\Yit =\Dit \theta + l_0(\vector X_{it}) + \alpha_i + \uit\\
   &\Dit = m_0(\vector X_{it}) + c_i + \vit \\
   & \alpha_i = 0.25 \,\bigg(\frac{1}{T} \sum_{t=1}^T\Dit-\overline{D}\Bigg) + 0.25 \,\frac{1}{T} \sum_{t=1}^T \sum_{k\in\mathcal{K}} X_{it,k} + a_i\\
   & a_i \distas{}{} N(0,0.95), \, X_{it,p} \distas{}{} N(0,5), \, c_i  \distas{}{} N(0,1).
\end{align}
\noindent where $k\in \mathcal{K}=\{1,3\}$ is the number of relevant (non-zero) confounding variables, and $p=30$ is the number of total confounding variables.

We consider three alternative functional forms of the confounders to model  the nuisance functions, $m_0$ and $l_0$, so as to vary in the level of non-linearity and non-smoothness of the functional forms.\footnote{We originally considered linear functional forms of the confounders but all estimators were performing as well as OLS.} \\

\noindent {\bf Design 1 (DGP1)}: Linear in the nuisance parameters
\begin{equation}
\begin{split}
  l_0(\vector X_{it}) &=  a \,X_{it,1} + X_{it,3} \\
  m_0(\vector X_{it}) &=  a\, X_{it,1} +X_{it,3} 
\end{split}
\end{equation}

\noindent  \text{\bf Design 2 (DGP2)}: Non-linear and smooth in the nuisance parameters
\begin{equation}
\begin{split}
 l_0(\vector X_{it}) &= \frac{\exp(X_{it,1})}{1+\exp(X_{it,1})} + a \,\cos(X_{it,3})\\
 m_0(\vector X_{it}) &= \cos(X_{it,1}) + a \,\frac{\exp(X_{it,3})}{1+\exp(X_{it,3})}
\end{split}
\end{equation}

\noindent  \text{\bf Design 3 (DGP3)}: Non-linear and discontinuous in the nuisance parameters
\begin{equation}
\begin{split}
l_0(\vector X_{it}) & = b\, (X_{it,1}\cdot X_{it,3}) + a \, (X_{it,3}\cdot\one[X_{it,3}>0])\\
m_0(\vector X_{it}) & =  a \, (X_{it,1}\cdot\one[X_{it,1}>0]) +b \, (X_{it,1}\cdot X_{it,3}),
\end{split}
\end{equation}
\noindent where $a=0.25$, $b=0.5$, $\one(z)=1$ if $z$ is true otherwise $\one(z)=0$. 

The population of cross-sectional units from which each Monte Carlo sub-sample is drawn consists of $N=1,000,000$ units observed over $T=10$ periods, but the study is conducting by subsampling $N=\{100, 1000, 4000\}$ to compare finite-sample performance for small, medium and large sample sizes. We run $R=100$ Monte Carlo replications per combination of sample size and learner. 

The nuisance parameters $l_0$ and $m_0$ are learned using LASSO, CART, gradient boosting and  RF; the hyperparameters for each algorithm are tuned via grid search, as described in Section~\ref{sec:tune_mc} of the Online Supplement.
LASSO uses an extended dictionary of the confounders, that is, a design matrix augmented with polynomials of order three and interaction terms between all the covariates. 
The hybrid and CRE approaches make use of twice the total number of the included covariates because their individual-specific means are included.

\subsection{Monte Carlo Simulation Results}\label{sec:simresults}
Tables~\ref{tab:appx_dgp1_collapse}-\ref{tab:appx_dgp3_collapse} display the simulation results in terms of average bias, RMSE, and the ratio of the standard error (SE) to the standard deviation (SD) for different cross-sectional sample sizes $N=\{100, 1000, 4000\}$ and fixed time $T=10$ across DGPs. Within each table, figures from different estimation approaches are shown. Different base learners (LASSO, CART, gradient boosting, and RF) used within DML are contrasted to OLS estimation.

\input{tabs/00_SI_simul_dgp1}
\input{tabs/00_SI_simul_dgp2}
\input{tabs/00_SI_simul_dgp3}

As expected, OLS yields the best performance under DGP1 (linear) in terms of bias, RMSE and SE/SD for all sample sizes such that an ensemble learning strategy (based on RMSE) would have chosen OLS over the ML algorithms. However, the bias and RMSE of the other estimators do decrease as the sample size increases, except for CART which performs worst on all measures for $N=1000$ and $N=4000$.
A similar pattern of results presents under DGP2 (nonlinear and smooth) except that CART and boosting estimators perform poorly in small samples. CART stands out from the other learners in having the poorest performance its estimated standard errors even in large samples. The robustness of OLS reflects that, despite being nonlinear, DGP2 is approximately linear over large areas of the $(x_1,x_3)$ support.
The pattern of results, however, changes under DGP3 (nonlinear and discontinuous) where the learners now outperform OLS in terms of RMSE and bias. 
OLS performs best in terms of its SE accuracy but is otherwise severely biased.  The tree-based learners perform well in terms of bias as the sample size increases, but under-estimate the standard errors with CART performing particular poorly.  An ensemble learning strategy would select the (extended-dictionary) LASSO even though it would lead to conservative inference with standard errors overestimated by about 20\%.

\begin{figure}[ht!]
    \centering
    \subfloat{\includegraphics[width= \textwidth]{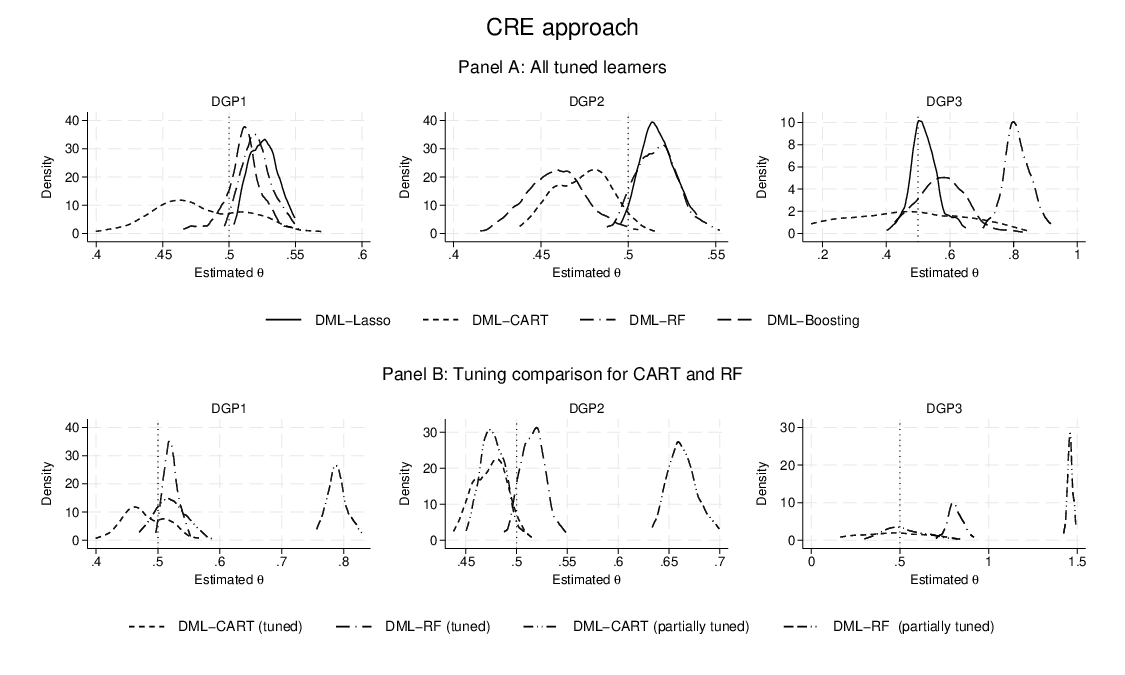}}
     \caption{Comparison of distributions of $\thetano$ across learners (Panel A) and with different tuned hyperparameters (Panel B) with the CRE approach.}
     \label{fig:kdensity_panels} 
\end{figure}

To understand more about the performance of the learners in terms of accuracy and precision, we inspect the sampling distributions of $\hat{\theta}_N$ with $N=1000$ for CRE in Panel A of Figure~\ref{fig:kdensity_panels}. 
Unlike the other learners, the sampling distribution of CART are seen to be highly non-normal for DGP3. 
However, this irregular behaviour is less severe under DGP1 and DGP2, where the estimated causal effects are close to normally distributed (but bimodal) and SE bias is slightly smaller. This indicates that statistical inference about $\theta_0$ can be unreliable using a single tree-based algorithms because the assumption of asymptotic normality does not hold.

One possible explanation for the non-normality of the sampling distributions is suboptimal hyperparameter tuning of the tree-based algorithms. The importance of optimal hyperparameter tuning for causal modelling has recently been shown for conditional average treatment effect estimators \citep{machlanski2023}. In particular, we were concerned that we had not tuned over a sufficiently wide range of values in the grid search or taken into account the adaptive nature of optimal hyperparameter choice where, for example, \citet[Theorem~1]{wager2015} have shown that the rate at which the minimum number of observations per leaf for `moderately high-dimensional' cases should increase with $N$ to control the error bounds on the resulting estimates. We hence experiment with an alternative tuning strategy for CART and RF to see whether we could find sampling distributions which were consistent and asymptotically normal but, in contrast to the results here, could only obtain asymptotically normal behaviour at the cost of substantial bias.

Panel B of Figure~\ref{fig:kdensity_panels} explores the possibility of suboptimal hyperparameter tuning by comparing the sampling distributions of $\thetano_N$ using the strategy from Section~\ref{sec:mc_design} (tuned) with an alternative tuning strategy (partially tuned) for CART and RF. The new tuning strategy for RF fixes the maximum depth of trees to 100, the number of trees in the forest to 1000, and select the optimal minimum node size between the interval $\{5,\lceil N/2 \rceil\}$. For CART, the maximum depth of the tree is set to 100, the optimal complexity parameter is selected from the interval $\{0.01, 0.05\}$, and the optimal minimum node size between the interval $\{5,\lceil N/2 \rceil\}$.
The new strategy forces each random tree to overfit the data and relies on a large forest to average out the overfitting errors. This leads to RF estimators with larger upward bias, but smaller SD. This is especially true for the non-linear discontinuous DGP3 where the new strategy leads to a clearly Gaussian normal sampling distribution. CART seems to benefit from this new tuning strategy displaying a unimodal sample distribution with smaller bias and SD in all DGP2 compared to the tuned counterpart. This may indicate that the single trees from the former CART replications underfit the data.

Overall, the new strategy is seen to be unsuccessful for RF, with the results for DGP3 indicating that the analyst must choose between bias and variance when using tree-based methods. 

%% file: tabs/00_SI_simul_dgp1.tex
\begin{table}[t!]
\centering
 \caption{Average MC simulation results for $\thetano_N$, DGP1 (linear)}\label{tab:appx_dgp1_collapse}
\scalebox{.8}{
\begin{threeparttable}	
\begin{tabular}{lccccccccc}  
\vspace{-3mm}\\   
\hline\hline
\vspace{-3mm}\\
&\multicolumn{3}{c}{$N=100$}&\multicolumn{3}{c}{$N=1,000$}&\multicolumn{3}{c}{$N=4,000$}\\  	
\cmidrule(l{.35cm}r{.25cm}){2-4}\cmidrule(l{.35cm}r{.25cm}){5-7}\cmidrule(l{.35cm}r{.25cm}){8-10}		
&	Bias&	RMSE	&	SE/SD	%
&	Bias&	RMSE	&	SE/SD	%
&	Bias&	RMSE	&	SE/SD	\\
\hline
\vspace{-1mm}\\
\multicolumn{10}{c}{\emph{Panel A: CRE approach}}\\
\vspace{-3mm}\\
OLS	&	0.003	&	0.033	&	1.036	&	0.001	&	0.010	&	1.057	&	0.001	&	0.005	&	1.123	\\
DML-Lasso	&	0.063	&	0.073	&	1.102	&	0.026	&	0.028	&	1.085	&	0.013	&	0.014	&	0.982	\\
DML-CART	&	-0.003	&	0.059	&	0.946	&	-0.020	&	0.040	&	0.462	&	-0.015	&	0.047	&	0.182	\\
DML-RF	&	0.090	&	0.103	&	1.054	&	0.021	&	0.024	&	1.231	&	0.011	&	0.012	&	1.137	\\
DML-Boosting	&	0.033	&	0.063	&	0.958	&	0.011	&	0.018	&	0.909	&	0.006	&	0.009	&	0.999	\\

\vspace{-1mm}\\
\multicolumn{10}{c}{\emph{Panel B: Exact approach with FD transformation}}\\
\vspace{-3mm}\\
OLS	&	0.003	&	0.041	&	1.002	&	0.000	&	0.012	&	1.016	&	0.001	&	0.006	&	1.029	\\
DML-Lasso	&	0.018	&	0.041	&	1.077	&	0.001	&	0.013	&	1.007	&	0.000	&	0.006	&	1.041	\\
DML-CART	&	0.041	&	0.065	&	0.792	&	0.031	&	0.052	&	0.303	&	0.064	&	0.072	&	0.190	\\
DML-RF	&	0.066	&	0.079	&	0.901	&	0.026	&	0.029	&	1.006	&	0.014	&	0.015	&	1.087	\\
DML-Boosting	&	-0.020	&	0.048	&	0.893	&	-0.013	&	0.021	&	0.793	&	-0.005	&	0.008	&	0.938	\\


\vspace{-1mm}\\
\multicolumn{10}{c}{\emph{Panel C: Approximation approach for WG transformation}}\\
\vspace{-3mm}\\
OLS	&	0.003	&	0.033	&	1.036	&	0.001	&	0.010	&	1.057	&	0.001	&	0.005	&	1.123	\\
DML-Lasso	&	0.009	&	0.032	&	1.081	&	0.002	&	0.010	&	1.073	&	0.001	&	0.005	&	1.008	\\
DML-CART	&	0.013	&	0.043	&	0.821	&	-0.005	&	0.027	&	0.408	&	0.018	&	0.032	&	0.201	\\
DML-RF	&	0.020	&	0.038	&	1.022	&	0.004	&	0.011	&	1.054	&	0.003	&	0.006	&	0.967	\\
DML-Boosting	&	-0.034	&	0.049	&	0.919	&	-0.007	&	0.014	&	0.853	&	-0.002	&	0.005	&	1.086	\\

\hline
\end{tabular} 	
\footnotesize
\renewcommand{\baselineskip}{11pt}
\textbf{Note:} The figures in the table are the average values over the total number of replications. The number of Monte Carlo replications is 100; the true target parameter is 0.50; $T=10$. We use the Neyman-orthogonal PO score and five-fold cross-fitting.
\end{threeparttable}
}
\end{table}

%% file: tabs/00_SI_simul_dgp2.tex
\begin{table}[t!]
\centering
 \caption{Average MC simulation results for $\thetano_N$, DGP2 (nonlinear and smooth)}\label{tab:appx_dgp2_collapse}
\scalebox{.8}{
\begin{threeparttable}	
\begin{tabular}{lccccccccc}  
\vspace{-3mm}\\   
\hline\hline
\vspace{-3mm}\\
&\multicolumn{3}{c}{$N=100$}&\multicolumn{3}{c}{$N=1,000$}&\multicolumn{3}{c}{$N=4,000$}\\  	
\cmidrule(l{.35cm}r{.25cm}){2-4}\cmidrule(l{.35cm}r{.25cm}){5-7}\cmidrule(l{.35cm}r{.25cm}){8-10}		
&	Bias&	RMSE	&	SE/SD	%
&	Bias&	RMSE	&	SE/SD	%
&	Bias&	RMSE	&	SE/SD	\\
\hline
\vspace{-1mm}\\
\multicolumn{10}{c}{\emph{Panel A: CRE approach}}\\
\vspace{-3mm}\\
OLS	&	0.001	&	0.029	&	1.003	&	0.000	&	0.008	&	1.044	&	0.000	&	0.004	&	1.063	\\
DML-Lasso	&	0.061	&	0.070	&	0.939	&	0.017	&	0.020	&	0.965	&	0.008	&	0.009	&	1.063	\\
DML-CART	&	-0.014	&	0.067	&	0.837	&	-0.026	&	0.031	&	0.846	&	-0.029	&	0.034	&	0.363	\\
DML-RF	&	0.019	&	0.042	&	1.013	&	0.017	&	0.020	&	1.000	&	0.005	&	0.008	&	0.987	\\
DML-Boosting	&	0.013	&	0.055	&	0.888	&	-0.041	&	0.044	&	0.848	&	0.002	&	0.008	&	0.910	\\

\vspace{-1mm}\\
\multicolumn{10}{c}{\emph{Panel B: Exact approach with FD transformation}}\\
\vspace{-3mm}\\
OLS	&	-0.001	&	0.037	&	0.932	&	-0.002	&	0.011	&	1.022	&	-0.001	&	0.005	&	1.021	\\
DML-Lasso	&	-0.002	&	0.035	&	0.969	&	-0.002	&	0.011	&	1.027	&	-0.001	&	0.005	&	1.018	\\
DML-CART	&	-0.042	&	0.054	&	0.980	&	-0.037	&	0.040	&	0.769	&	-0.039	&	0.043	&	0.313	\\
DML-RF	&	-0.020	&	0.040	&	0.979	&	0.001	&	0.012	&	1.026	&	-0.003	&	0.007	&	1.018	\\
DML-Boosting	&	-0.040	&	0.059	&	0.839	&	-0.009	&	0.018	&	0.794	&	-0.001	&	0.007	&	0.916	\\


\vspace{-1mm}\\
\multicolumn{10}{c}{\emph{Panel C: Approximation approach for WG transformation}}\\
\vspace{-3mm}\\
OLS	&	0.001	&	0.032	&	0.993	&	0.002	&	0.008	&	1.199	&	0.001	&	0.004	&	1.156	\\
DML-Lasso	&	0.006	&	0.030	&	1.118	&	0.001	&	0.009	&	1.089	&	0.001	&	0.005	&	0.988	\\
DML-CART	&	-0.013	&	0.036	&	0.945	&	-0.009	&	0.015	&	0.833	&	-0.005	&	0.009	&	0.747	\\
DML-RF	&	-0.004	&	0.033	&	0.981	&	-0.002	&	0.010	&	1.054	&	0.000	&	0.005	&	0.984	\\
DML-Boosting	&	-0.042	&	0.055	&	0.889	&	-0.008	&	0.014	&	0.847	&	-0.001	&	0.005	&	1.129	\\

\hline
\end{tabular} 	
\footnotesize
\renewcommand{\baselineskip}{11pt}
\textbf{Note:} The figures in the table are the average values over the total number of replications. The number of Monte Carlo replications is 100; the true target parameter is 0.50; $T=10$. We use the Neyman-orthogonal PO score and five-fold cross-fitting.
\end{threeparttable}
}
\end{table}

%% file: tabs/00_SI_simul_dgp3.tex
\begin{table}[t!]
\centering
 \caption{Average MC simulation results for $\thetano_N$, DGP3 (nonlinear and discontinuous)}\label{tab:appx_dgp3_collapse}
\scalebox{.8}{
\begin{threeparttable}	
\begin{tabular}{lccccccccc}  
\vspace{-3mm}\\   
\hline\hline
\vspace{-3mm}\\
&\multicolumn{3}{c}{$N=100$}&\multicolumn{3}{c}{$N=1,000$}&\multicolumn{3}{c}{$N=4,000$}\\  	
\cmidrule(l{.35cm}r{.25cm}){2-4}\cmidrule(l{.35cm}r{.25cm}){5-7}\cmidrule(l{.35cm}r{.25cm}){8-10}		
&	Bias&	RMSE	&	SE/SD	%
&	Bias&	RMSE	&	SE/SD	%
&	Bias&	RMSE	&	SE/SD	\\
\hline
\vspace{-1mm}\\
\multicolumn{10}{c}{\emph{Panel A: CRE approach}}\\
\vspace{-3mm}\\
OLS	&	0.993	&	0.993	&	0.891	&	0.993	&	0.993	&	0.910	&	0.993	&	0.993	&	0.999	\\
DML-Lasso	&	0.103	&	0.130	&	1.062	&	0.021	&	0.049	&	0.778	&	0.009	&	0.014	&	1.235	\\
DML-CART	&	0.136	&	0.250	&	0.405	&	-0.032	&	0.177	&	0.187	&	-0.087	&	0.199	&	0.084	\\
DML-RF	&	0.698	&	0.711	&	0.438	&	0.312	&	0.315	&	1.349	&	0.149	&	0.151	&	1.320	\\
DML-Boosting	&	0.406	&	0.447	&	0.603	&	0.077	&	0.108	&	0.731	&	-0.007	&	0.033	&	0.871	\\

\vspace{-1mm}\\
\multicolumn{10}{c}{\emph{Panel B: Exact approach with FD transformation}}\\
\vspace{-3mm}\\
OLS	&	0.993	&	0.993	&	0.889	&	0.993	&	0.993	&	0.961	&	0.993	&	0.993	&	0.951	\\
DML-Lasso	&	0.036	&	0.051	&	1.143	&	0.004	&	0.013	&	1.045	&	0.005	&	0.008	&	1.050	\\
DML-CART	&	0.615	&	0.634	&	0.341	&	0.316	&	0.409	&	0.091	&	0.291	&	0.385	&	0.049	\\
DML-RF	&	0.965	&	0.966	&	0.415	&	0.846	&	0.850	&	0.146	&	0.752	&	0.765	&	0.071	\\
DML-Boosting	&	0.718	&	0.723	&	0.557	&	0.269	&	0.284	&	0.506	&	-0.014	&	0.059	&	0.461	\\


\vspace{-1mm}\\
\multicolumn{10}{c}{\emph{Panel C: Approximation approach for WG transformation}}\\
\vspace{-3mm}\\
OLS	&	0.993	&	0.993	&	0.891	&	0.993	&	0.993	&	0.910	&	0.993	&	0.993	&	0.999	\\
DML-Lasso	&	0.973	&	0.973	&	0.939	&	0.977	&	0.977	&	1.002	&	0.977	&	0.977	&	1.120	\\
DML-CART	&	0.731	&	0.737	&	0.479	&	0.724	&	0.735	&	0.100	&	0.754	&	0.764	&	0.050	\\
DML-RF	&	0.962	&	0.962	&	0.520	&	0.969	&	0.969	&	0.841	&	0.972	&	0.972	&	0.739	\\
DML-Boosting	&	0.817	&	0.818	&	0.666	&	0.879	&	0.879	&	0.506	&	0.918	&	0.918	&	0.426	\\

\hline
\end{tabular} 	
\footnotesize
\renewcommand{\baselineskip}{11pt}
\textbf{Note:} The figures in the table are the average values over the total number of replications. The number of Monte Carlo replications is 100; the true target parameter is 0.50; $T=10$. We use the Neyman-orthogonal PO score and five-fold cross-fitting.
\end{threeparttable}
}
\end{table}

%% file: SI_tuning.tex
\section{Hyperparameter Tuning}\label{sec:hyperpara}
\subsection{The Grid Search Algorithm}\label{sec:grid_search}
Finding the optimal configuration of hyperparameters (or hyperparameter tuning) of a learner is essential to reach state-of-the-art performance in effect estimation, independently of the choice of estimators and learners, whereas relying on default hyperparameter values (suggested by statistical packages or the literature) severely compromises the ability of learners to reach their full potential, ultimately biasing the causal estimand as shown in \citet{machlanski2023, machlanski2024} and \citet{bach2024hyper}. 

Hyperparameter optimization proceeds with trials of different configurations of values of the hyperparameters to tune. Resampling methods -- such as, cross-validation (CV) -- are used to evaluate the performance of the algorithm in terms of RMSE (when the hyperparametwers are numeric). This procedure is repeated for several configurations until a stopping rule is applied (e.g., maximum number of evaluations). Finally, the configuration with the best performance (with, e.g.,  lowest RMSE) is selected and passed to the learner to train and test the model. 

In the DML algorithm based on \citet{bach2024}, hyperparameter tuning works as follows. 
\setcounter{bean}{0}
\begin{center}
\begin{list}
{\textsc{Step} \arabic{bean}.}{\usecounter{bean}}
    \item When the tuning is on folds,  units in the training sample for fold $k$ ($ W^c_k$) are used for tuning. These are subsequently divided, e.g., in five-fold CV to create training and testing inner samples. When tuning is not on folds (default), all data is passed to the tuning procedure, but the composition of the units assigned to the $k$-th CV fold differs from the corresponding fold in the DML procedure. Then, five-fold CV is instantiated such that the $k$-th CV fold is the test sample and the rest the training sample. 
    \item The models are tuned with grid search, which is a hyperparameter optimizer algorithm that evaluates the performance of base learners with different configurations of the hyperparameter values \citep{bergstra2012}. The optimizer randomly searches among a specified number of different values to try per hyperparameter (or \texttt{resolution}) and stops the optimization when the specified maximum number of evaluations is reached  (or \texttt{terminator}).
    This method are non-adaptive such that the proposed configuration ignores the performance of previous ones.

    \item Each evaluation within the tuning routine selects the best configuration of hyperparameters among all $k$ CV folds, based on the lowest RMSE. Once the tuning algorithm stops (e.g., at the $j$-th evaluation), the best configuration of hyperparameters among the $j$ results is chosen (based on lower RMSE) and passed to the DML algorithm.

    \item The best configuration is set as parameters of the learners of the nuisance parameters. The model is then trained on the complementary set for fold $k$, $W^c_k$, and tested on $W_k$. Predictions for $m$ and $l$ are stored.
\end{list}
\end{center}

\noindent The default tuning procedure for DML (not on folds) follows the same sample splitting principle behind DML. There is no separate test set for validation because predictions are done at the DML stage, and the test sample in the learning stage of DML uses different combinations of units in each fold (tuning not on folds).

\subsection{Hyperparameter Tuning in Monte Carlo Simulations}\label{sec:tune_mc}
The hyperparameters of base learners used in Monte Carlo simulation exercises described in Section~\ref{sec:mcsimul} and in the Online Supplement~\ref{sec:simresults} are tuned as follows.
LASSO uses the penalization parameter equivalent to minimum mean cross-validated error. The hyperparameters of tree-based  are tuned via grid search \citep{bergstra2012}.  We set the hyperparameter optimizer to try five distinct values per hyperparameter, randomly selected from the intervals specified in Table~\ref{tab:hyperpara}, within each evaluation and terminate the optimization at the fifth evaluation. 

 \input{tabs/tab_design}

CART's hyperparameters that are tuned are the complexity parameter, minimum number of observations in terminal node, maximum depth of final tree. For gradient-boosting, the hyperparameters that are tuned are the L2 regularization term on weights, and the maximum depth of final trees; the number of decision trees in the final model is fixed to 100. The hyperparameters that are tuned for RF are the number of trees, minimum number of observations in terminal node, maximum depth of any node of the final tree; the number of covariates randomly sampled to split at each node is set to the maximum.\footnote{We do not randomly sub-select the predictors to grow the random forest using either the default values ($p/3$ for regression problems; $\sqrt{p}$ for classification problems) or tuning the parameters $M$, \texttt{mtry}, because we are concerned that  \mbox{\ref{item:asm_selection_fe}} (ignorable selection) would collapse if few predictors are included. Because our Monte Carlo simulation design involves only two relevant predictors while the rest is noise, setting \texttt{mtry} to the largest value possible ensures that these two variables will always be selected for splitting and, hence, contribute to the prediction. This is common practice with datasets with few relevant variables, such as genetic datasets \citep{probst2019}.}

\subsection{Hyperparameter Tuning in the Empirical Application}\label{sec:tune_appl}
The hyperparameters of CART, RF, and gradient boosting  in Section~\ref{sec:application} are tuned via grid search with the criteria as follows. The hyperparameter optimization algorithm evaluates 10 different combinations of values of the hyperparameters to tune (\texttt{n\_evals}) and terminate at the twentieth round (\texttt{resolution}); then, the best combination in terms of RMSE performance is selected by the algorithm and used for learning the functional form of the nuisance parameters. 

For CART, the complexity parameter (\texttt{cp}) is tuned by randomly searching between 0.001 and 0.05, the minimum number of observations in terminal node (\texttt{minbucket}) as default, and maximum depth of final tree (\texttt{maxdepth}) between 2 and 10. For gradient-boosting,  the number of decision trees in the final model (\texttt{nrounds}) is set to 1000, the L2 regularization term on weights (\texttt{lambda}) is tuned by selecting an optimal value between 0 a and 2, and the maximum depth of final trees (\texttt{max\_depth}) between 2 and 10. Regarding RF, the number of trees (\texttt{num.trees}) is set to 1000, the number of covariates randomly sampled to split at each node (\texttt{mtry}) is set to be the total number of covariates (no sub-sampling), the minimum number of observations in terminal node (\texttt{min.node.size}) is as per default, the maximum depth (\texttt{max.depth}) is tuned between 2 and 10. L1-penalization parameter for LASSO is relative to the model with minimum mean cross-validated error (\texttt{lambda.min}). 

%% file: tabs/tab_design.tex
\begin{table}[h!]
\centering
 \caption{Hyperparameter tuning in MC simulations}\label{tab:hyperpara}
\scalebox{.68}{
\begin{threeparttable}	
   \begin{tabular}{lllll}  
\hline\hline
Learner&Hyperparamter& Value of parameter in set & Description\\	
\hline
\vspace{-2mm}\\
Lasso&\texttt{lambda.min}& cross-validated&$\lambda$ equivalent to minimum mean cross-validated error\\
\vspace{-2mm}\\

CART&\texttt{cp}&real value in \{0.001,0.05\}&Prune all nodes with a complexity less than cp from the printout.\\
&\texttt{minbucket} & default&Minimum number of observations in any terminal <leaf> node.\\
&\texttt{maxdepth}&integer in \{2,10\}&Maximum depth of any node of the final tree.\\

\vspace{-2mm}\\
Boosting&\texttt{lambda}&real value in \{0,5\}&L2 regularization term on weights.\\
&\texttt{maxdepth}&integer in \{2,10\}&Maximum depth of any node of the final tree.\\
&\texttt{nrounds}& 100&Number of decision trees in the final model\\

\vspace{-2mm}\\
RF& \texttt{num.trees}&100 &Number of trees in the forest.\\
&\texttt{min.node.size}&default&Minimal node size to split at. \\
&\texttt{max.depth}&integer in \{2,10\}&Maximum depth of any node of the final tree.\\
&\texttt{mtry}& all covariates &The number of covariates, randomly sampled, to split at each node.\\
\vspace{-2mm}\\
\hline
\end{tabular} 
\footnotesize
\renewcommand{\baselineskip}{11pt}
\textbf{Note:}  The hyperparameters of the base learners chosen to model the nuisance functions are tuned in each Monte Carlo replication via grid search, that evaluates each possible combination of hyperparameters' values in the grid  \citep{bergstra2012}. We set the hyperparameter optimizer to try ten distinct values per hyperparameter randomly selected from the specified intervals, within each evaluation and terminate the optimization at the twentieth evaluation.
\end{threeparttable}}
\end{table}


%% file: SI_var_description.tex
\section{Variable Description}\label{sec:appx_var_des}
Table~\ref{tab:appx_descriptive} reports the means, minimum and maximum values of all the variables used in the empirical analysis. The confounding variables are divided if four groups:  \emph{base control variables}, which include the five variables  that are used in the original study (with the exception of age square); \emph{socio-economic variables};  \emph{work-related variables}; and \emph{ideology-related variables}.

\input{tabs/00_SI_descriptives}

%% file: tabs/00_SI_descriptives.tex
\begin{table}[th!]
\centering
 \caption{Descriptive statistics}\label{tab:appx_descriptive}
\scalebox{.9}{
\begin{threeparttable}	
\begin{tabular}{llll}  
\vspace{-3mm}\\
\hline\hline
            &        Mean&         Min&         Max\\
\hline
\vspace{-2mm}\\
\multicolumn{4}{l}{\emph{Dependent variable}}\\
Vote for conservative parties&        0.25&        0&        1\\

\vspace{-2mm}\\
\multicolumn{4}{l}{\emph{Regressor}}\\
NMW         &        0.02&           0&           1\\

\vspace{-2mm}\\
\multicolumn{4}{l}{\emph{Base control variables}}\\
Household size      &        3.05&           1&          14\\
Household income (in log)   &        8.10&          -1.35&          12.31\\
Age         &       40.58&          18&          65\\
Degree    &        0.16&           0&           1\\
Married    &        0.64&           0&           1\\

\vspace{-2mm}\\
\multicolumn{4}{l}{\emph{Socio-economic variables}}\\
Woman      &        0.00&           0&           0\\
White ethnic background    &        0.08&           0&           1\\
Years of education     &       12.76&           6&          18\\
Age left school&       16.21&           8&          24\\
No. children     &        0.67&           0&           7\\
No. time married   &        0.77&           0&           3\\
Monthly family credit&      176.23&           0&       15184.50\\
Recipient of family credit &        0.32&           0&           1\\

\vspace{-2mm}\\
\multicolumn{4}{l}{\emph{Work-related variables}}\\
Days unemployment&        8.88&           0&        8130\\
Days in self-employment&       11.56&           0&       13514\\
Days in employment&     2343.06&           0&       18554\\
Days in inactivity&       67.80&           0&       14002\\
No. spells in wave&        1.13&           1&          11\\
No. spells in employment in wave&        1.00&           1&           7\\
No. spells in unemployment in wave&        0.02&           0&           3\\
No. spells in inactivity in wave&        0.04&           0&           9\\
No. spells in self-employment in wave&        0.01&           0&           4\\
Work experience  &            24.37&        0&       55\\
In maternity leave&        0.00&           0&           0\\ 
Full-time job&        0.78&           0&           1\\
Manager&        0.22&           0&           1\\
Average firm size &      232.09&           0&        1000\\
In small firm &        0.60&           0&           1\\
Worked Hours     &       37.52&           0&         161\\
In private sector   &        0.62&           0&           1\\

\vspace{-2mm}\\
\multicolumn{4}{l}{\emph{Ideology-related variables}}\\
Feels more British&        0.03&           0&           1\\
Thinks homosexuality is wrong&        0.06&           0&           1\\
Thinks family suffers from woman in full-time work&        0.15&           0&           1\\
In a trade union    &        0.31&           0&           1\\
Member of a religious group&        0.06&           0&           1\\

Hours spent in house work   &        9.70&           0&          99\\
\vspace{-2mm}\\
\hline
No. Observations       &       59,745&            &            \\
Number of clusters &   9,922 &            &            \\
\hline
\end{tabular} 	
\footnotesize
\renewcommand{\baselineskip}{11pt}
\textbf{Note: The sample of respondents spans from wave 1 until wave 18.} 
\end{threeparttable}
}
\end{table}

%% file: main_arxiv.bbl
\begin{thebibliography}{}

\bibitem[\protect\citeauthoryear{Athey and Imbens}{Athey and
  Imbens}{2016}]{atheyimbens2016}
Athey, S. and G.~Imbens (2016).
\newblock Recursive partitioning for heterogeneous causal effects.
\newblock {\em Proceedings of the National Academy of Sciences\/}~{\em
  113\/}(27), 7353--7360.

\bibitem[\protect\citeauthoryear{Athey, Tibshirani, and Wager}{Athey
  et~al.}{2019}]{atheytibshirani2019}
Athey, S., J.~Tibshirani, and S.~Wager (2019).
\newblock Generalized random forests.
\newblock {\em The Annals of Statistics\/}~{\em 47\/}(2), 1148 -- 1178.

\bibitem[\protect\citeauthoryear{Bach, Chernozhukov, Kurz, Spindler, and
  Klaassen}{Bach et~al.}{2024}]{bach2024}
Bach, P., V.~Chernozhukov, M.~S. Kurz, M.~Spindler, and S.~Klaassen (2024).
\newblock {DoubleML} -- {A}n object-oriented implementation of double machine
  learning in {R}.
\newblock {\em Journal of Statistical Software\/}~{\em 108\/}(3), 1--56.

\bibitem[\protect\citeauthoryear{Bach, Chernozhukov, and Spindler}{Bach
  et~al.}{2023}]{bach2023}
Bach, P., V.~Chernozhukov, and M.~Spindler (2023).
\newblock {Heterogeneity in the US gender wage gap}.
\newblock {\em Journal of the Royal Statistical Society Series A: Statistics in
  Society\/}~{\em 187\/}(1), 209--230.

\bibitem[\protect\citeauthoryear{Bach, Schacht, Chernozhukov, Klaassen, and
  Spindler}{Bach et~al.}{2024}]{bach2024hyper}
Bach, P., O.~Schacht, V.~Chernozhukov, S.~Klaassen, and M.~Spindler (2024,
  01--03 Apr).
\newblock Hyperparameter tuning for causal inference with double machine
  learning: A simulation study.
\newblock In {\em Proceedings of the Third Conference on Causal Learning and
  Reasoning}, Volume 236 of {\em Proceedings of Machine Learning Research},
  pp.\  1065--1117. PMLR.

\bibitem[\protect\citeauthoryear{Baiardi and Naghi}{Baiardi and
  Naghi}{2024a}]{baiardi2024plough}
Baiardi, A. and A.~A. Naghi (2024a).
\newblock The effect of plough agriculture on gender roles: A machine learning
  approach.
\newblock {\em Journal of Applied Econometrics\/}.

\bibitem[\protect\citeauthoryear{Baiardi and Naghi}{Baiardi and
  Naghi}{2024b}]{baiardi2024}
Baiardi, A. and A.~A. Naghi (2024b).
\newblock The value added of machine learning to causal inference: Evidence
  from revisited studies.
\newblock {\em The Econometrics Journal\/}, utae004.

\bibitem[\protect\citeauthoryear{Belloni, Chernozhukov, and Hansen}{Belloni
  et~al.}{2014}]{belloni2014}
Belloni, A., V.~Chernozhukov, and C.~Hansen (2014).
\newblock High-dimensional methods and inference on structural and treatment
  effects.
\newblock {\em Journal of Economic Perspectives\/}~{\em 28\/}(2), 29--50.

\bibitem[\protect\citeauthoryear{Belloni, Chernozhukov, Hansen, and
  Kozbur}{Belloni et~al.}{2016}]{belloni2016}
Belloni, A., V.~Chernozhukov, C.~Hansen, and D.~Kozbur (2016).
\newblock Inference in high-dimensional panel models with an application to gun
  control.
\newblock {\em Journal of Business and Economic Statistics\/}~{\em 34\/}(4),
  590--605.

\bibitem[\protect\citeauthoryear{Bergstra and Bengio}{Bergstra and
  Bengio}{2012}]{bergstra2012}
Bergstra, J. and Y.~Bengio (2012).
\newblock Random search for hyper-parameter optimization.
\newblock {\em Journal of Machine Learning Research\/}~{\em 13\/}(2), 281--305.

\bibitem[\protect\citeauthoryear{Bia, Huber, and Laff{\'e}rs}{Bia
  et~al.}{2023}]{huber2023}
Bia, M., M.~Huber, and L.~Laff{\'e}rs (2023).
\newblock Double machine learning for sample selection models.
\newblock {\em Journal of Business and Economic Statistics\/}, 1--12.

\bibitem[\protect\citeauthoryear{Breiman}{Breiman}{1996}]{breiman1996}
Breiman, L. (1996).
\newblock Stacked regressions.
\newblock {\em Machine learning\/}~{\em 24}, 49--64.

\bibitem[\protect\citeauthoryear{Chang}{Chang}{2020}]{chang2020}
Chang, N.-C. (2020).
\newblock Double/debiased machine learning for difference-in-differences
  models.
\newblock {\em The Econometrics Journal\/}~{\em 23\/}(2), 177--191.

\bibitem[\protect\citeauthoryear{Chernozhukov, Chetverikov, Demirer, Duflo,
  Hansen, Newey, and Robins}{Chernozhukov et~al.}{2018}]{chernozhukov2018}
Chernozhukov, V., D.~Chetverikov, M.~Demirer, E.~Duflo, C.~Hansen, W.~Newey,
  and J.~Robins (2018).
\newblock Double/debiased machine learning for treatment and structural
  parameters.
\newblock {\em The Econometrics Journal\/}~{\em 21\/}(1),
  \textsc{C}1--\textsc{C}68.

\bibitem[\protect\citeauthoryear{Chernozhukov, Newey, and Singh}{Chernozhukov
  et~al.}{2022}]{chernozhukov2022}
Chernozhukov, V., W.~K. Newey, and R.~Singh (2022).
\newblock Automatic debiased machine learning of causal and structural effects.
\newblock {\em Econometrica\/}~{\em 90\/}(3), 967--1027.

\bibitem[\protect\citeauthoryear{Di~Francesco}{Di~Francesco}{2024}]{difrancesco2022}
Di~Francesco, R. (2024).
\newblock Aggregation trees.
\newblock {\em arXiv preprint arXiv:2410.11408\/}.
\newblock First version; submitted on 15 Oct 2024.

\bibitem[\protect\citeauthoryear{Fazio and Reggiani}{Fazio and
  Reggiani}{2023}]{fazio2023}
Fazio, A. and T.~Reggiani (2023).
\newblock Minimum wage and tolerance for high incomes.
\newblock {\em European Economic Review\/}~{\em 155}, 104445.

\bibitem[\protect\citeauthoryear{Klosin and Vilgalys}{Klosin and
  Vilgalys}{2023}]{klosin2022}
Klosin, S. and M.~Vilgalys (2023).
\newblock Estimating continuous treatment effects in panel data using machine
  learning with an agricultural application.
\newblock {\em arXiv preprint arXiv:2207.08789\/}.
\newblock Second version; last revised 13 Sep 2023.

\bibitem[\protect\citeauthoryear{Knaus}{Knaus}{2022}]{knaus2022double}
Knaus, M.~C. (2022).
\newblock Double machine learning-based programme evaluation under
  unconfoundedness.
\newblock {\em The Econometrics Journal\/}~{\em 25\/}(3), 602--627.

\bibitem[\protect\citeauthoryear{Kock}{Kock}{2016}]{kock2016}
Kock, A.~B. (2016).
\newblock Oracle inequalities, variable selection and uniform inference in
  high-dimensional correlated random effects panel data models.
\newblock {\em Journal of Econometrics\/}~{\em 195\/}(1), 71--85.

\bibitem[\protect\citeauthoryear{Kock and Tang}{Kock and Tang}{2019}]{kock2019}
Kock, A.~B. and H.~Tang (2019).
\newblock Uniform inference in high-dimensional dynamic panel data models with
  approximately sparse fixed effects.
\newblock {\em Econometric Theory\/}~{\em 35\/}(2), 295--359.

\bibitem[\protect\citeauthoryear{Koles{\'a}r, M{\"u}ller, and
  Roelsgaard}{Koles{\'a}r et~al.}{2023}]{kolesar2023}
Koles{\'a}r, M., U.~K. M{\"u}ller, and S.~T. Roelsgaard (2023).
\newblock The fragility of sparsity.
\newblock {\em arXiv preprint arXiv:2311.02299\/}.

\bibitem[\protect\citeauthoryear{Lechner}{Lechner}{2015}]{lechner2015}
Lechner, M. (2015).
\newblock Treatment effects and panel data.
\newblock In {\em The Oxford Handbook of Panel Data (Online Edition)}. Oxford
  Academic.

\bibitem[\protect\citeauthoryear{Lechner and Mareckova}{Lechner and
  Mareckova}{2022}]{lechner2022}
Lechner, M. and J.~Mareckova (2022).
\newblock Modified causal forest.
\newblock {\em arXiv preprint arXiv:2209.03744\/}.

\bibitem[\protect\citeauthoryear{Machlanski, Samothrakis, and
  Clarke}{Machlanski et~al.}{2023}]{machlanski2023}
Machlanski, D., S.~Samothrakis, and P.~Clarke (2023).
\newblock Hyperparameter tuning and model evaluation in causal effect
  estimation.
\newblock {\em arXiv preprint arXiv:2303.01412\/}.

\bibitem[\protect\citeauthoryear{Machlanski, Samothrakis, and
  Clarke}{Machlanski et~al.}{2024}]{machlanski2024}
Machlanski, D., S.~Samothrakis, and P.~S. Clarke (2024, 01--03 Apr).
\newblock Robustness of algorithms for causal structure learning to
  hyperparameter choice.
\newblock In {\em Proceedings of the Third Conference on Causal Learning and
  Reasoning}, Volume 236 of {\em Proceedings of Machine Learning Research},
  pp.\  703--739. PMLR.

\bibitem[\protect\citeauthoryear{Mundlak}{Mundlak}{1978}]{mundlak1978}
Mundlak, Y. (1978).
\newblock On the pooling of time series and cross section data.
\newblock {\em Econometrica\/}~{\em 46\/}(1), 69--85.

\bibitem[\protect\citeauthoryear{Probst, Wright, and Boulesteix}{Probst
  et~al.}{2019}]{probst2019}
Probst, P., M.~N. Wright, and A.-L. Boulesteix (2019).
\newblock Hyperparameters and tuning strategies for random forest.
\newblock {\em Wiley Interdisciplinary Reviews: Data mining and knowledge
  discovery\/}~{\em 9\/}(3), e1301.

\bibitem[\protect\citeauthoryear{Robinson}{Robinson}{1988}]{robinson1988}
Robinson, P.~M. (1988).
\newblock Root-n-consistent semiparametric regression.
\newblock {\em Econometrica: Journal of the Econometric Society\/}~{\em
  56\/}(4), 931--954.

\bibitem[\protect\citeauthoryear{Rubin}{Rubin}{1974}]{rubin1974}
Rubin, D.~B. (1974).
\newblock Estimating causal effects of treatments in randomized and
  nonrandomized studies.
\newblock {\em Journal of Educational Psychology\/}~{\em 66\/}(5), 688--701.

\bibitem[\protect\citeauthoryear{Sela and Simonoff}{Sela and
  Simonoff}{2012}]{sela2012}
Sela, R.~J. and J.~S. Simonoff (2012).
\newblock Re-em trees:\ a data mining approach for longitudinal and clustered
  data.
\newblock {\em Machine Learning\/}~{\em 86}, 169--207.

\bibitem[\protect\citeauthoryear{Semenova, Goldman, Chernozhukov, and
  Taddy}{Semenova et~al.}{2023}]{semenova2023}
Semenova, V., M.~Goldman, V.~Chernozhukov, and M.~Taddy (2023).
\newblock Inference on heterogeneous treatment effects in high-dimensional
  dynamic panels under weak dependence.
\newblock {\em Quantitative Economics\/}~{\em 14\/}(2), 471--510.

\bibitem[\protect\citeauthoryear{Strittmatter}{Strittmatter}{2023}]{strittmatter2023}
Strittmatter, A. (2023).
\newblock What is the value added by using causal machine learning methods in a
  welfare experiment evaluation?
\newblock {\em Labour Economics\/}~{\em 84}, 102412.

\bibitem[\protect\citeauthoryear{{University of Essex, Institute for Social and
  Economic Research}}{{University of Essex, Institute for Social and Economic
  Research}}{2018}]{bhps}
{University of Essex, Institute for Social and Economic Research} (2018).
\newblock British \textsc{H}ousehold \textsc{P}anel \textsc{S}urvey:
  \textsc{W}aves 1-18, 1991-2009.
\newblock [data collection]. 8th Edition. UK Data Service.
\newblock SN: 5151, DOI: http://doi.org/10.5255/UKDA-SN-5151-2.

\bibitem[\protect\citeauthoryear{van~der Laan, Polley, and Hubbard}{van~der
  Laan et~al.}{2007}]{laan2007}
van~der Laan, M.~J., E.~C. Polley, and A.~E. Hubbard (2007).
\newblock Super learner.
\newblock {\em Berkeley Devision of Biostatistics Working Paper Series\/}, 222,
  pp.1--20.

\bibitem[\protect\citeauthoryear{Wager and Athey}{Wager and
  Athey}{2018}]{atheywager2018}
Wager, S. and S.~Athey (2018).
\newblock Estimation and inference of heterogeneous treatment effects using
  random forests.
\newblock {\em Journal of the American Statistical Association\/}~{\em
  113\/}(523), 1228--1242.

\bibitem[\protect\citeauthoryear{Wager and Walther}{Wager and
  Walther}{2015}]{wager2015}
Wager, S. and G.~Walther (2015).
\newblock Adaptive concentration of regression trees, with application to
  random forests.
\newblock {\em arXiv preprint arXiv:1503.06388\/}.

\bibitem[\protect\citeauthoryear{Wooldridge}{Wooldridge}{2019}]{wooldridge2019}
Wooldridge, J.~M. (2019).
\newblock Correlated random effects models with unbalanced panels.
\newblock {\em Journal of Econometrics\/}~{\em 211\/}(1), 137--150.

\bibitem[\protect\citeauthoryear{Wooldridge and Zhu}{Wooldridge and
  Zhu}{2020}]{wooldridge2020}
Wooldridge, J.~M. and Y.~Zhu (2020).
\newblock Inference in approximately sparse correlated random effects probit
  models with panel data.
\newblock {\em Journal of Business and Economic Statistics\/}~{\em 38\/}(1),
  1--18.

\end{thebibliography}
